\documentclass[preprint,floatfix,aps,prd,showpacs,nofootinbib,amsmath,amssymb,amsfonts,superscriptaddress]{revtex4-2}
\usepackage[utf8]{inputenc}
\usepackage{bm}
\usepackage{amsmath,amssymb}
\usepackage{graphicx}
\usepackage{physics}
\usepackage{dsfont}
\usepackage[normalem]{ulem}
\usepackage{hyperref,url}
\usepackage{mathrsfs}
\usepackage{calrsfs}
\usepackage{tikz}
 \usepackage{comment}

\usepackage[caption=false]{subfig}

\usepackage[active]{srcltx}

\expandafter\ifx\csname package@font\endcsname\relax\else
\expandafter\expandafter
\expandafter\usepackage
\expandafter\expandafter
\expandafter{\csname package@font\endcsname}%
\fi
\hypersetup{
	colorlinks=true,       
	linkcolor=blue,          
	citecolor=blue,        
}

\usepackage[section]{placeins}

\usepackage{fancybox}
\usepackage{soul}

\begin{document}
\title{Generation of effective massive Spin-2 fields through spontaneous symmetry breaking of scalar field}

\author{Susobhan Mandal}
\email{sm12ms085@gmail.com}

\author{S. Shankaranarayanan}
\email{shanki@iitb.ac.in}

\affiliation{ Department of Physics, 
Indian Institute of Technology Bombay,
Mumbai - 400076, India }

\date{\today}

\begin{abstract}
General relativity and quantum field theory are the cornerstones of our understanding of physical processes, from subatomic to cosmic scales. While both theories work remarkably well in their tested domains, they show minimal overlap. However, our research challenges this separation by revealing that non-perturbative effects bridge these distinct domains. We introduce a novel mechanism wherein, at linear order, spin-2 fields around an arbitrary background acquire \emph{effective mass} due to the spontaneous symmetry breaking (SSB) of either global or local symmetry of complex scalar field minimally coupled to gravity. The action of the spin-2 field is identical to the extended Fierz-Pauli (FP) action, corresponding to the mass deformation parameter $\alpha = 1/2$.
We show that this occurs due to the effect of SSB on the variation of the energy-momentum tensor of the matter field, which has a dominant effect during SSB. The extended FP action has a salient feature, compared to the standard FP action: the action has 6 degrees of freedom with no ghosts. For local $U(1)$ SSB, we establish that the effective mass of spin-2 fields is related to the mass of the gauge boson and the electric charge of the complex scalar field. Interestingly, our results indicate that the millicharged dark matter scalar fields, generating dark photons, can produce a mass of spin-2 fields of the same order as the Hubble constant $(H_0)$. Hence, we argue that the dark sector offers a natural explanation for the acceleration of the current Universe.
\end{abstract}

\maketitle
\newpage

\section{Introduction}

Einstein's general relativity (GR) and quantum field theory represent two successful frameworks in modern physics, each applicable within its respective domain. However, the amalgamation of these theories often leads to observational discrepancies. One prominent example is the cosmological constant (CC), typically interpreted as the energy contribution from vacuum fluctuations of matter fields~\cite{Zeldovich:1968ehl, Weinberg:1988cp}. It is challenge because all the attempts that try to explain the $10^{120}$ orders of magnitude of discrepancy between the theory and observation have produced unsatisfying results~\cite{Weinberg:1988cp,Carroll:2000fy, Padmanabhan:2002ji,Nobbenhuis:2004wn,mannheim2011comprehensive,Martin:2012bt,donoghue2021cosmological}. 

Various astrophysical and cosmological observations, like Type Ia Supernova, Baryon Acoustic Oscillation and Cosmic 
microwave background, provide compelling evidence for the accelerating universe~\cite{SupernovaCosmologyProject:1998vns,SupernovaSearchTeam:1998fmf,ESSENCE:2007acn,SupernovaCosmologyProject:2008ojh,WMAP:2012nax,SDSS:2014iwm,BOSS:2016wmc,Planck:2018vyg}. In the standard cosmological model, this phenomenon is explained by introducing \emph{dark energy}, a component that interacts with the rest of the Universe solely through gravity and possesses significant negative pressure, driving the Universe's acceleration~\cite{Peebles:2002gy,Copeland:2006wr,Frieman:2008sn,Caldwell:2009ix,Silvestri:2009hh,
2016-Joyce.etal-ARN,Huterer:2017buf,Kamionkowski:2022pkx}.  CC is a viable candidate for the dark energy~\cite{2021-DiValentino.etal-Class.Quant.Grav.,2022-Schoeneberg.etal-Phys.Rept.}.
Unfortunately, the nature of dark energy remains unknown, leading to a large number of hypotheses and candidates for its role as a fundamental component of the Universe's energy budget~\cite{Peebles:2002gy,Copeland:2006wr,Frieman:2008sn,Caldwell:2009ix,Silvestri:2009hh,
2016-Joyce.etal-ARN,Huterer:2017buf,Kamionkowski:2022pkx}.

An alternative approach is to attribute the current acceleration due to 
the differing nature of gravity on the cosmological scales~\cite{Sotiriou:2008rp,Koyama:2015vza,Nojiri:2017ncd,Olmo:2019flu,2022-Shanki.Joseph-GRG}. Combined with the fact that GR is not tested with high-precision outside solar system~\cite{Willbook}, a well-crafted modified gravity theory holds the potential to address the CC problem and late-time acceleration of the Universe ~\cite{Sotiriou:2008rp,Koyama:2015vza,Nojiri:2017ncd,Olmo:2019flu,2022-Shanki.Joseph-GRG}. One of the critical requirements of any modified gravity theory is that it must meet the stringent constraints imposed by the solar system experiments~\cite{Willbook}. 

Massive gravity, an extension of GR, is one of the simplest modified gravity models whose departure from GR is governed by a single parameter — the graviton mass $(m)$~\cite{de2014massive,deRham:2013qqa,dubovsky2004phases,deRham:2010ik, de2011helicity,de2011resummation,  hassan2011non, hinterbichler2012theoretical,de2012ghost, de2013massive, de2014massive}. The Vainshtein screening mechanism ensures the recovery of GR in $m \rightarrow 0$ limit \cite{Babichev:2013usa, Gumrukcuoglu:2021gua}. From an observational viewpoint, massive gravity can be easily made to be consistent with most tests of GR (effectively placing an upper bound on $m$) without spoiling its role in the cosmological scales~\cite{Gershtein:1997mm, gershtein2004graviton, choudhury2004probing, gruzinov2005graviton, brito2013massive,zakharov2016constraining, gupta2018limit, rana2018bounds, will2018solar, desai2018limit, miao2019bounding, de2021minimal, bernus2020constraint, shao2020new, bernus2019constraining, Goldhaber:1974wg, Goldhaber:2008xy, Talmadge:1988qz, Finn:2001qi}.  For instance, taking $m$ to be comparable to the Hubble constant $H_{0}$ can lead to late-time acceleration~\cite{deRham:2013qqa,Gershtein:1997mm}.

The effect of mass of spin-2 fields on the cosmological scales is analogous to the propagation of electromagnetic (EM) waves in plasma. In unmagnetized neutral plasma, the dispersion relation for EM waves is $\omega^2 = k^2 + \omega_p^2$, where $\omega_p$ is plasma frequency. Thus, waves with frequencies lower than $\omega _{p}$ cannot propagate in plasma. Similarly, in the case of massive gravity, high frequencies propagate in the background space-time without any attenuation. However, the low-frequency waves are attenuated, leading to a phenomenon known as \emph{degravitation} \cite{deRham:2014zqa, deRham:2007rw, Dvali:2007kt} (see Appendix \ref{appendix: degravitation}). As the CC is inherently a constant, it operates like a source with an exceedingly long wavelength \cite{ROVELLI20071287}. Hence, it acts as a screen that serves as a high-pass filter, leading to a considerably diminished observed effective value \cite{Dvali:2007kt}. 

However, massive gravity theories \emph{do not} naturally explain the origin of the mass of spin-2 fields, leading to considerable skepticism about this approach~\cite{comelli2013massive, bellazzini2018beyond, pitts2007universally}. 
Also, the standard Fierz-Pauli action~\cite{Gambuti:2021meo} corresponding to the massive spin-2 field in arbitrary curved space-time has a ghost~\cite{Bengtsson:1994vn} (see the discussion in Appendix \ref{Appendix D}). Recently, it was shown that there exists a nonlinear completion of the Fierz-Pauli action that is free of ghost up to the fourth order~\cite{Hassan:2011vm, Golovnev:2011aa, Izumi:2013poa}). Our work addresses this gap by providing a natural explanation for the generation of \emph{effective mass of spin-2 fields} through spontaneous symmetry breaking (SSB) of a $U(1)$ complex scalar field. We also show that the extended Fierz-Pauli action does not have ghost DoF and, hence, is stable in the linear regime.

SSB is a unifying concept that explains, among other things, ferromagnetism, superconductivity, Bose-Einstein condensation, and the Higgs mechanism~\cite{Coleman:1985rnk,beekman2019introduction,kapusta1981bose, coleman1974spontaneous, PhysRevD.42.3587, CONSOLI1985653}. SSB arises from the interplay between nonlinearity and symmetric potentials. It is well known that the symmetry of the ground state in models with nonlinear interactions mirrors the symmetry of the external/self-interacting potential \cite{beekman2019introduction}. This is always true in quantum mechanics (and any linear theory) as long as the nonlinearity remains sufficiently weak. However, SSB occurs when the nonlinearity reaches a critical strength. This phenomenon is easily observed in double-well potentials \cite{beekman2019introduction}. SSB plays a pivotal role in the standard model of particle physics, particularly about the Higgs~\cite{weinberg1973perturbative, weinberg1976implications, miransky2002spontaneous,friederich2013gauge,kibble2015spontaneous}.

We demonstrate that SSB of a scalar field can lead to non-trivial infrared (IR) effects on gravity. By using the complex scalar field minimally coupled to gravity, we present a unique approach for the origin of mass of spin-2 fields. The effective mass of the spin-2 fields is related to the mass of the scalar field and the self-interaction coupling constant. In the case of local U(1) SSB, we show that the mass of spin-2 fields is related to the mass of the gauge boson and the electric charge of the complex scalar field. Our results show that the millicharged dark matter scalar fields that generate dark photons can lead to graviton mass of the same order as the Hubble constant $(H_0)$. Consequently, the mechanism provides a \emph{natural explanation} for the current acceleration of the universe. In the remainder of this work, we explicitly demonstrate how the mass of spin-2 fields arises from the non-zero VEV of the complex scalar field in three different scenarios: Global $U(1)$ invariance, local  $U(1)$ invariance and Coleman-Weinberg mechanism~\cite{Coleman:1973jx,Jackiw:1974cv}.

The rest of this work is organized as follows: In Sec.~\eqref{sec:setup}, we discuss the setup and define the fourth-order tensor ${\mathcal{O}_{\mu\nu}^{ \ \ \rho\sigma}}$. As a first application of our work,  in Sec.~\eqref{sec:GlobalSSB},  we obtain the effective mass for the spin-2 field via global U(1) SSB. In Sec.~\eqref{sec:LocalSSB}, we delve deeper into our analysis and demonstrate that effective massive spin-2 is generated via local U(1) SSB. Sec.~\eqref {sec:Discussion} discusses the key results and possible implications in the early Universe cosmology. Appendices \eqref{Appendix.B.new} - \eqref{appendix: degravitation} provide details of the calculations in the main text. In this work, we set $ \hbar = c = 1$ and $ \kappa^ {2} = 8\pi G$.

\section{The Setup}
\label{sec:setup}

To keep things transparent, we start with a complex scalar field ($\Phi$) minimally coupled to gravity in a general 4-D spacetime. The action is:
\begin{equation}\label{eq. M6}
S = \frac{1}{2\kappa^{2}} \int \!\! d^{4}x \sqrt{-g} R 
+ S_{\Phi}(\Phi, \partial_\mu \Phi) \, .
\end{equation}
The action of a global $U(1)$ invariant complex scalar field is:
\begin{equation}\label{eq. M1}
S_{\Phi} = - \int d^{4}x\sqrt{-g}\Big[g^{\mu\nu}\partial_{\mu}\Phi\partial_{\nu}\Phi^{*}
 + U(|\Phi|)\Big] \, ,
\end{equation}
where $U(|\Phi|)$ is self-interacting potential. Varying the above action \eqref{eq. M6} w.r.t the metric leads to Einstein's equations ($G_{\mu\nu} = \kappa^2\,  T_{\mu\nu}$), 
where the energy-momentum tensor $(T_{\mu\nu})$ is:
\begin{eqnarray}
\label{eq. M3}
T_{\mu\nu}  = - \frac{2}{\sqrt{-g}}\frac{\delta S_{\Phi}}{\delta g^{\mu\nu}} & =& \left[
\partial_{\mu}\Phi\partial_{\nu}\Phi^{*} + \partial_{\nu}\Phi\partial_{\mu}\Phi^{*} \right. \\ 
 & - &
 \nonumber
\left. g_{\mu\nu}\left(g^{\rho\sigma}\partial_{\rho}\Phi\partial_{\sigma}\Phi^{*} 
 + U(|\Phi|)\right)\right] \, .
\end{eqnarray}
For a generic complex scalar field $\Phi(t,\bf{x})$, there are ten components of the energy-momentum tensor $T_{\mu\nu}$. On the other hand, the conservation of the energy-momentum tensor leads to 6 independent components of the energy-momentum tensor. In a similar manner, $\nabla^{\mu}G_{\mu\nu} = 0$ leads to 6 independent components of Einstein's tensor in general. As a result, the number of independent components for a generic complex scalar field remains the same on both sides of Einstein's field equations. This is consistent with the number of independent metric components in arbitrary 4-D spacetime being 6. It is easy to verify that the conservation of the stress-tensor $\nabla^{\mu}T_{\mu\nu} = 0$ leads to the equations of motion. We now define a fourth-rank mixed tensor which is the metric variation of $T_{\mu\nu}$:
\begin{eqnarray}\label{eq. M4}
{\mathcal{O}_{\mu\nu}^{ \ \ \rho\sigma}} &=& \frac{\delta T_{\mu\nu}(x)}{\delta 
g_{\rho\sigma}(y)} - \frac{1}{2}g^{\rho\sigma}T_{\mu\nu}\delta^{4}(x - y) \nonumber  \\
& = & \Bigg[- \frac{1}{2}\left(\delta_{\mu}^{\rho}\delta_{\nu}^{\sigma} + \delta_{\mu}
^{\sigma}\delta_{\nu}^{\rho}\right)\Big[g^{\alpha\beta}\partial_{\alpha}\Phi\partial
_{\beta}\Phi^{*} + U(|\Phi|)\Big] \\
& + & \frac{1}{2}g_{\mu\nu}(\partial^{\rho}\Phi\partial^{\sigma}\Phi^{*}
 + \partial^{\sigma}\Phi\partial^{\rho}\Phi^{*})\Bigg]\delta^{(4)}(x - y) \nonumber \\
 & - & \frac{1}{2}g^{\rho\sigma}T_{\mu\nu}\delta^{4}(x - y) \nonumber 
\end{eqnarray}
Note that we can also define a related fourth-rank covariant tensor ${\mathcal{O}_{\mu\nu\rho\sigma}}(x - y)$ which is obtained by the variation of the stress-tensor w.r.t the inverse metric. In this work, we will use the mixed tensor defined above. It is important to note that, like $T_{\mu\nu}$, the fourth rank ${\mathcal{O}_{\mu\nu}^{ \ \ \rho\sigma}}$ is defined for any arbitrary space-time. Note that the presence of the Dirac delta function ensures that the operator is local, ultimately resulting in a local action for the spin-2 field.

Einstein's field equations are highly non-linear, which restricts the possibility of finding solutions to only a few highly symmetric cases. The perturbative approach is essential to understand the nature of the exact solutions beyond the symmetric cases. However, to conduct a perturbative expansion, it is necessary to establish a background --- a manifold representing the average (Universe) or background space-time. The process of averaging in GR is ambiguous due to the involvement of complex non-linear interactions. The influence of these non-linearities on the overall expansion is referred to as backreaction, which remains significant even at late times \cite{Buchert:2007ik, Buchert:2011sx}.

In our approach, we will refrain from making any assumptions about the background metric ($\bar{g}_{\mu\nu}$) for two reasons: First, we aim to capture the influence of the non-linear potential on perturbations, particularly those arising from spontaneous symmetry breaking (SSB). In other words, we want to examine the contribution of the SSB potential to the background itself. Therefore, our analysis requires a non-flat background metric, denoted by $\bar{g}_{\mu\nu}$. Second, the potential \( U(|\Phi|) \) that we will consider is highly non-linear, and because the scalar field is dynamic, it is not possible to impose any specific symmetry on the background metric.

As we show in the rest of this work, despite the unknown exact form of the background metric, our approach adeptly captures the dynamics of SSB on the spin-2 fields. Based on this, we now consider the following metric decomposition:
\begin{equation}\label{metric decomposition 1}
g_{\mu\nu} = \bar{g}_{\mu\nu} + 2\kappa h_{\mu\nu}, 
~~g^{\mu\nu} \simeq \bar{g}^{\mu\nu} - 2\kappa h^{\mu\nu} + 4\kappa^{2}h^{\mu\rho} h_{ \ \rho}^{\nu},
\end{equation}
where $\bar{g}_{\mu\nu}$ is the metric describing the background geometry \textit{w.r.t} which we linearize the action \eqref{eq. M6}, $h_{\mu\nu}$ is the metric perturbation and $\kappa$ carries the order of the metric perturbation\footnote{By linearizing, we refer to the decomposition presented in the above equation where $2\kappa |h_{\mu\nu}| \ll |\bar{g}_{\mu\nu}|$. This essentially implies that $h_{\mu\nu}$ can be treated as a metric perturbation with respect to the background metric $\bar{g}_{\mu\nu}$, which satisfies the background Einstein field equations.}.
Since the background metric is arbitrary, it has 6 independent components. Hence, the metric perturbations $h_{\mu\nu}$ also have six independent components.

Expanding the action \eqref{eq. M6} at the quadratic level of metric fluctuation $h_{\mu\nu}$ (details in Appendix~\ref{Appendix.B.new}), we obtain:
{
\begin{eqnarray}
& & S_{h} = \int d^{4}x\sqrt{-\bar{g}}\Big[ - \frac{1}{2}\bar{\nabla}_{\lambda}h_{\mu\nu}\bar{\nabla}
 ^{\lambda}h^{\mu\nu} + \frac{1}{2}\bar{\nabla}_{\rho}h\bar{\nabla}^{\rho}h \nonumber\\
 & + & \bar{\nabla}^{\nu}h^{\rho\mu}\bar{\nabla}_{\mu}h_{\rho\nu} - \bar{\nabla}_{\mu}h\bar{\nabla}_{\nu}h^{\mu\nu} - \frac{\bar{R}}{2}\left[h_{\alpha\beta}h^{\alpha\beta} - \frac{h^{2}}{2}\right]\nonumber\\
 & + & 2\bar{R}_{\nu\sigma}\left[h^{\nu\lambda}h_{ \ \lambda}^{\sigma} - \frac{1}{2}h
 h^{\nu\sigma}\right] \Bigg] \nonumber \\
\label{eq. M8}
& - & \kappa^{2}\int d^{4}x\sqrt{-\bar{g}} \, h^{\mu\nu} \, 
\overline{\mathcal{O}_{\mu\nu}^{ \ \ \rho\sigma}} \, h_{\rho\sigma},
\end{eqnarray}
}
where
\begin{eqnarray}
& & \overline{\mathcal{O}_{\mu\nu}^{ \ \ \rho\sigma}}  = - \frac{1}{2}\Bigg[\left(\delta_{\mu}^{\rho}\delta_{\nu}^{\sigma} + \delta_{\mu}^{\sigma}\delta_{\nu}^{\rho}\right)
\Big[\bar{g}^{\alpha\beta}\partial_{\alpha}\Phi\partial_{\beta}\Phi^{*} + U(|\bar{\Phi}|)\Big] \nonumber \\
&& ~~~- \frac{1}{2} \bar{g}_{\mu\nu}(\partial^{\rho}\Phi\partial^{\sigma}\Phi^{*}
 + \partial^{\sigma}\Phi\partial^{\rho}\Phi^{*})\Bigg] 
- \frac{1}{2} \bar{g}^{\rho\sigma}\Big[\partial_{\mu}\Phi\partial_{\nu}\Phi^{*} 
 \nonumber \\
\label{eq. M10}
&& ~~~+ \partial_{\nu}\Phi\partial_{\mu}\Phi^{*} - \bar{g}_{\mu\nu}\left(\bar{g}^{\alpha\beta}\partial_{\alpha}\Phi\partial_{\beta}
 \Phi^{*} + U(|\bar{\Phi}|)\right)\Big] \, , 
\end{eqnarray}
where $\overline{\mathcal{O}_{\mu\nu}^{ \ \ \rho\sigma}}$ is the fourth-rank tensor \eqref{eq. M4} 
obtained w.r.t. the arbitrary background metric $\bar{g}_{\mu\nu}$. 
The above action \eqref{eq. M8} is a key expression regarding which we would like to discuss the following: The last term contains the information about the variation of the stress-tensor as a function of metric. In cosmology all quantities vary with time; this is captured via Ricci scalar/tensor. At thermal equilibrium,  these terms dominate over the last term. This will be explicitly shown in Sec.~\eqref{sec:LocalSSB}, where we consider a local U(1) SSB of a complex scalar field.
However, the contribution of the last term is significant when the Universe is \emph{out-of-equilibrium}, like at SSB or phase transition~\cite{Garbrecht:2018mrp, Kolb:1983ni}.

As mentioned earlier, GR is highly non-linear, and obtaining exact solutions is possible only for specific symmetric space-times. Given this, perturbation theory in GR can be approached in two ways: The first approach does not assume any symmetry of the background metric and only requires that it be a solution to Einstein's field equations. Alternatively, one can assume a specific form for the background metric and utilize its symmetries to derive the perturbation equations. The second approach is more widely used in black hole physics and cosmology~\cite{1992-Mukhanov.etal-PRep,2008-Tsagas.etal-PRep,1983-Chandrasekhar-MathematicalTheoryBlack}.

In this work, we use the first approach and the fourth-rank tensor \eqref{eq. M4} contributes to the backreaction of the stress-energy tensor onto the background space-time. The second approach involves extracting the constant $U_{0}$ from the symmetry-breaking field potential  $U(|\Phi|)$ at the tree-level action. The contribution to the mass of spin-2 field coming from the field $ \Phi $ after SSB will vanish since we have already extracted the non-zero minima value $U_{0}$ from the field potential. 

\section{Effective massive spin-2 fields through global $U(1)$ SSB}
\label{sec:GlobalSSB}
To go about understanding the effect of SSB on metric perturbations, we consider the following potential, invariant under the global $U(1)$ symmetry \cite{kibble2015spontaneous, friederich2013gauge}:
\begin{equation}\label{eq. M5}
U(|\Phi|) = - m_{\Phi}^{2}|\Phi|^{2} + \frac{\lambda}{2}|\Phi|^{4}.
\end{equation}
where $m_{\Phi}$ has mass dimension $1$ and $\lambda$ is the dimensionless self-interaction coupling constant. For $m_{\Phi}^{2} > 0$ this symmetry is spontaneously broken. Hence, the field $|\Phi|$ assumes a non-zero vacuum expectation value $\phi_{0}$:
\begin{equation}\label{eq. M12}
\phi_{0} = {m_{\Phi}}/{\sqrt{\lambda}},
\end{equation} 
leading to the following value of the field potential:
\begin{equation}\label{eq. M13}
U(\phi_{0}) = - {m_{\Phi}^{4}}/{(2\lambda)}.
\end{equation}
Substituting $U(\phi_0)$ in action \eqref{eq. M8} and using the Einstein's equations for the background geometry, $\bar{R} = 
- \kappa^{2} \bar{T}$ and $\bar{R}_{\nu\sigma} = \kappa^{2}\left(\bar{T}_{\nu\sigma} - \frac{1}{2}\bar{g}
_{\nu\sigma} \bar{T} \right)$ where $\bar{T} = - 4 U(\phi_{0})$ and $\bar{T}_{\nu\sigma} = - \bar{g}_{\nu\sigma}
U(\phi_{0})$, leads to (details in Appendix \ref{Appendix.B.new}):
{\small
\begin{eqnarray}
\label{eq. M14}
& & S_{h}  = \int d^{4}x\sqrt{-\bar{g}}\Big[ - \frac{1}{2}\bar{\nabla}_{\lambda}h_{\mu\nu}\bar{\nabla}
 ^{\lambda}h^{\mu\nu} + \frac{1}{2}\bar{\nabla}_{\rho}h\bar{\nabla}^{\rho}h  \nonumber\\
&+& \!\! \bar{\nabla}^{\nu}
 h^{\rho\mu}\bar{\nabla}_{\mu}h_{\rho\nu} - \bar{\nabla}_{\mu}h\bar{\nabla}_{\nu}h^{\mu\nu}
- \frac{1}{2}m^{2}(h_{\alpha\beta}h^{\alpha\beta} - \alpha h^{2})\Big] \, 
\end{eqnarray}
}
where,
\begin{equation}\label{eq. M15}
m = {\kappa m_{\Phi}^{2}}/({2\sqrt{\lambda}}) \, , \, \alpha = 1/2 \, .
\end{equation}
This is the first key result of this work, and we want to discuss the following points. First, the above action corresponds to a massive spin-2 field in arbitrary curved spacetime and is known as the \emph{generalized Fierz-Pauli action with $\alpha$ being massive deformation parameter} (see Appendix \ref{appendix C} for more details). It is important to note that the Fierz-Pauli action is a special case of the above action where $\alpha = 1$. 

Second, while the Fierz-Pauli action in Minkowski spacetime has only five dynamical polarizations, it has six in arbitrary curved spacetime. The additional polarization is a ghost \cite{Bengtsson:1994vn} (see the discussion in Appendix \ref{Appendix D}). As can be seen from Eq.~(\ref{eq. M8}), the Fierz-Pauli theory does not have any ghost DOF in Einstein spaces, defined by the relation $R_{\mu\nu} = \Lambda g_{\mu\nu}$. However, this can not extended to arbitrary background as the action depends on the background Ricci tensor and Ricci scalar~\cite{Buchbinder:1999ar, Bernard:2014bfa, Bernard:2015uic,Mazuet:2017hey, Mazuet_2018}. Recently, nonlinear extension of massive gravity gains much attention in order to take into account the IR effect within GR~(see, for instance, \cite{Hassan:2011vm, Golovnev:2011aa, Izumi:2013poa}). Here, as shown in Appendices \ref{Appendix D} and \ref{appendix: degravitation}, the extended Fierz-Pauli action does not have ghost DoF and hence is stable in the linear regime.

The flat spacetime limit of the above action reduces to (details in Appendix \ref{section 3}):
\begin{eqnarray}
\label{eq. M14A}
S_{h} & = & \int d^{4}x \Big[ - \frac{1}{2}\partial_{\lambda}h_{\mu\nu}\partial^{\lambda}h^{\mu\nu} + \frac{1}{2}\partial_{\rho}h\partial^{\rho}h\\
 &+& \partial^{\nu}h^{\rho\mu}\partial_{\mu}h_{\rho\nu} - \partial_{\mu}h\partial_{\nu}h^{\mu\nu} - \frac{1}{2}m^{2}(h_{\mu\nu}h^{\mu\nu} - \alpha h^{2})\Big].  \nonumber 
\end{eqnarray}     
The above action is identical to the \emph{extended Fierz-Pauli (FP) action} corresponding to massive deformation parameter $\alpha = 1/2$~\cite{Comelli:2012vz,Crisostomi:2014shq}. (See Appendix \ref{appendix C} for more details.) The above action differs from the FP action, corresponding to $\alpha = 1$~\cite{deRham:2014zqa}. For $\alpha = 1$, $h$ is traceless, and $h_{\mu\nu}$ has five degrees of freedom (DoF). However, for $\alpha = {1}/{2}$, $h$ satisfies the wave equation: 
\begin{equation}
(\Box - m^{2})h = 0. 
\end{equation}
Thus, for $\alpha = 1/2$, $h_{\mu\nu}$ has six DoF.  Although naively, this seems to be a ghost-like DoF \cite{Creminelli:2005qk}, $\alpha = {1}/{2}$ has no ghost DoF~\cite{Comelli:2012vz,Crisostomi:2014shq}. 
In Appendices \ref{appendix C} and \ref{Appendix D}, we have explicitly shown that the extended Fierz-Pauli action has no ghosts. Hence, the theory is well-defined in arbitrary curved spacetime. It is important here to note that since the background metric $\bar{g}_{\mu\nu}$ is chosen to be arbitrary, it has six independent components, which also implies that the resultant metric perturbation $h_{\mu\nu}$ has 6 DoF which is mathematically consistent with our analysis. On the other hand, it is also important to note that in asymptotic spatial infinity, the DoF of metric perturbation reduces to 2 as the energy-momentum vanishes there identically. As a result, there would not be any SSB in the matter sector.
Thus, we conclude that the global U(1) SSB of a minimally coupled, self-interacting complex scalar field \emph{leads to an effective mass of spin-2 fields} which is proportional to the mass of the scalar field ($m_\Phi$) and inversely proportional to self-interacting coupling ($\lambda$).

Third, since SSB occurs at a specific VEV, the kinetic term of the scalar field vanishes, and only the value of the potential at SSB contributes to the effective mass of spin-2 fields. Fourth, it is essential to note that although the global $U(1)$ symmetry breaking leads to an effective mass of spin-2 fields, it also produces a Goldstone Boson, which would produce a long-range attractive potential. Since this is inevitable and its existence is questionable due to various observational constraints~\cite{Kibble:2015mwa}, this SSB phenomenon might be debatable. To overcome this, we must look for a scenario where massless modes do not occur. This naturally leads us to generalize 
our analysis to local $U(1)$ invariant theories.

Lastly, as previously noted, the non-linear nature of Einstein's field equations limits the availability of exact solutions, making a perturbative approach essential. Although the averaging process in GR is ambiguous due to complex non-linear interactions, our method effectively captures the influence of the non-linear potential on perturbations, particularly as it relates to SSB. The traditional perturbative approach used in black hole physics and cosmology~\cite{1992-Mukhanov.etal-PRep,2008-Tsagas.etal-PRep,1983-Chandrasekhar-MathematicalTheoryBlack} typically assumes a background geometry \emph{ab initio}. 
As mentioned earlier, our approach is different from the standard perturbation theory. The standard perturbation theory involves isolating the constant $U_{0}$ from the field potential at the tree-level action, followed by decomposing the metric into a background component and perturbations. This ultimately yields the same expression for the mass of the tensor perturbations as obtained earlier in this section. However, this approach carries a limitation: the background geometry is assumed to be either de Sitter or anti-de Sitter, solely based on $U_{0}$. In contrast, the approach used in this work does not make any specific assumptions about the background metric, other than that it satisfies the background Einstein field equations. It is particularly effective in capturing the influence of the non-linear potential due to SSB, even when the exact form of the background metric remains undetermined.

\section{Effective massive spin-2 fields through local U(1) SSB}
\label{sec:LocalSSB}

Here, we consider complex scalar field ($\Phi$) with local $U(1)$ symmetry, with the following matter Lagrangian:
\begin{equation}
{\cal L}_{M} =  - \frac{1}{4} F_{\mu\nu} F_{\mu\nu} - g^{\mu\nu}\mathcal{D}_{\mu}\Phi\mathcal{D}_{\nu}\Phi^{*} - U(|\Phi|) ,
\end{equation}
where $F_{\mu\nu} = \partial_{\mu}A_{\nu} - \partial
_{\nu}A_{\mu}$ is the field tensor of the $U(1)$ gauge field ($A_{\mu}$) and $\mathcal{D}_{\mu} = \partial_{\mu} - i g_{_A} A_{\mu}$ where $g_{_A}$ is the gauge coupling constant of the $U(1)$ group~\cite{Coleman:1973jx}. The energy-momentum tensor corresponding to the above matter action is
\begin{eqnarray}
\bar{T}_{\mu\nu} & =&  \mathcal{D}_{\mu}\bar{\Phi}\mathcal{D}_{\nu}\bar{\Phi}^{*} + \mathcal{D}_{\nu}\bar{\Phi}\mathcal{D}_{\mu}\bar{\Phi}^{*} + \bar{F}_{\mu\alpha}\bar{F}_{\nu\beta}\bar{g}^{\alpha\beta}\\
&-& \frac{1}{4}\bar{g}_{\mu\nu}\bar{F}_{\rho\sigma}\bar{F}_{\alpha\beta}\bar{g}^{\rho\alpha}\bar{g}^{\beta\sigma} - \bar{g}_{\mu\nu}[\bar{g}^{\rho\sigma}\mathcal{D}_{\rho}\bar{\Phi}\mathcal{D}_{\sigma}\bar{\Phi}^{*} + U(|\bar{\Phi}|)]. \nonumber
\end{eqnarray}
As a result, $\overline{\mathcal{O}_{\mu\nu}^{ \ \ \rho\sigma}}$ evaluated at the background metric  is:
\begin{eqnarray}
\overline{\mathcal{O}_{\mu\nu}^{ \ \ \rho\sigma}} & =& - \frac{1}{2}\bar{F}_{\mu\alpha}\bar{F}_{\nu\beta}(\bar{g}^{\alpha\rho}\bar{g}^{\beta\sigma} + \bar{g}^{\alpha\sigma}\bar{g}^{\beta\rho}) - \frac{1}{8}(\delta_{\mu}^{\rho}\delta_{\nu}^{\sigma} + \delta_{\mu}^{\sigma}\delta_{\nu}^{\rho}) \nonumber\\
 & \times& \bar{F}_{\lambda\kappa}\bar{F}_{\alpha\beta}\bar{g}^{\lambda\alpha}\bar{g}^{\kappa\beta} + \frac{1}{8}
 \bar{g}_{\mu\nu}\bar{F}_{\lambda\kappa}\bar{F}_{\alpha\beta}(\bar{g}^{\lambda\rho}\bar{g}^{\alpha\sigma} + \bar{g}^{\lambda
 \sigma}\bar{g}^{\alpha\rho})\nonumber\\
 & \times&  \bar{g}^{\beta\kappa} + \frac{1}{8}\bar{g}_{\mu\nu}\bar{F}_{\lambda\kappa}\bar{F}_{\alpha\beta}\bar{g}^{\lambda\alpha}(\bar{g}^{\beta\rho}\bar{g}^{\kappa\sigma} + \bar{g}^{\beta\sigma}\bar{g}^{\kappa\rho})\\
 & -& \frac{1}{2}(\delta_{\mu}^{\rho}\delta_{\nu}^{\sigma} + \delta_{\mu}^{\sigma}\delta_{\nu}
 ^{\rho})[\bar{g}^{\alpha\beta}\mathcal{D}_{\alpha}\bar{\Phi}\mathcal{D}_{\beta}\bar{\Phi}^{*} + U(|\bar{\Phi}|)]\nonumber\\
 & +& \frac{1}{2}\bar{g}_{\mu\nu}(\mathcal{D}^{\rho}\bar{\Phi}\mathcal{D}^{\sigma}\bar{\Phi}^{*} + \mathcal{D}^{\sigma}\bar{\Phi}\mathcal{D}^{\rho}\bar{\Phi}^{*}) - \frac{1}{2}\bar{g}^{\rho\sigma}\bar{T}_{\mu\nu}\nonumber.
\end{eqnarray}
Like earlier, we choose $U(|\Phi|)$ to be a symmetry breaking potential \eqref{eq. M5} where $m_{\Phi} > 0$. 
%
%
Thus, after the symmetry breaking $|\Phi|$, this leads to the same value of $\phi_{0}$ given in Eq.~\eqref{eq. M12} and 
$U(\phi_{0})$ given by Eq.~\eqref{eq. M13}. Because of the symmetry breaking the $U(1)$ gauge field also acquires mass whose value is 
$m_{A} = g_{_A} \, m_{\Phi} \sqrt{2/\lambda}$. Substituting these values, using the Einstein's equations for the background geometry, and neglecting the gauge and complex field fluctuations at SSB, we obtain 
\begin{equation}
\label{eq:OSSB}
\overline{\mathcal{O}_{\mu\nu}^{ \ \ \rho\sigma}} \sim \frac{m_{\Phi}^{4}}{4\lambda}(\delta_{\mu}^{\rho}\delta_{\nu}^{\sigma} + \delta_{\mu}^{\sigma}\delta_{\nu}^{\rho}
 - g_{\mu\nu}g^{\rho\sigma}).
\end{equation}
Substituting this in action \eqref{eq. M8} leads to:
{\small 
\begin{eqnarray}
S_{h} & = & \int d^{4}x\sqrt{-\bar{g}}\Big[ - \frac{1}{2}\bar{\nabla}_{\lambda}h_{\mu\nu}\bar{\nabla}
 ^{\lambda}h^{\mu\nu} + \frac{1}{2}\bar{\nabla}_{\rho}h\bar{\nabla}^{\rho}h  \\
 & + & \bar{\nabla}^{\nu}h^{\rho\mu}\bar{\nabla}_{\mu}h_{\rho\nu} - \bar{\nabla}_{\mu}h\bar{\nabla}_{\nu}h^{\mu\nu}
 - \kappa^{2}\frac{m_{\Phi}^{4}}{2\lambda}\left[h_{\alpha\beta}h^{\alpha\beta} - \frac{h^{2}}{2}\right]\Big]. \nonumber
\end{eqnarray}
}
This is the second key result of this work, and we want to discuss the following points. First, like in the previous case, the above action corresponds to \emph{extended FP action} corresponding to the mass deformation parameter $\alpha = 1/2$. Like in the previous case, the trace of the metric tensor is dynamical, and the metric tensor has six DoF with no ghost (details in Appendices  \ref{Appendix D} and \ref{appendix: degravitation}). 

Second, by rewriting the mass of spin-2 fields in terms of the mass of the gauge field ($m_A$), we have:
\begin{equation}
\label{eq:GravMass-LocalU1}
m = \frac{\kappa}{4}\frac{m_{A}^{2}}{g_{_A}^{2}}\sqrt{\lambda} \, . 
\end{equation}
Thus, the experimental bound on $m_A$ can directly provide a bound on graviton mass $m$. Till now, the analysis has been general without any assumption on the choice of the gauge field. 

Let us now take the gauge field to be EM. Using the upper bound on the mass of the photons to be $10^{-18}$eV~\cite{Ryutov:2007zz}, $g_{_A}$ to be EM gauge coupling constant, and $\lambda < 1$, the upper bound on the graviton mass is 
\[
m < 2.72 \times 10^{-64}~{\rm eV}  \, .
\]
which is consistent with the bounds put by LIGO~\cite{abbott2021tests, LIGOScientific:2021sio} and other cosmological observations~\cite{Gershtein:1997mm, gershtein2004graviton, choudhury2004probing, gruzinov2005graviton, brito2013massive,
deRham:2013qqa, zakharov2016constraining, gupta2018limit, rana2018bounds, will2018solar, desai2018limit, miao2019bounding, de2021minimal, bernus2020constraint, shao2020new, bernus2019constraining, Goldhaber:1974wg, Goldhaber:2008xy, Talmadge:1988qz, Finn:2001qi}. However, the value of $m$ is small to provide a natural explanation for the acceleration of the current Universe.

This brings us to the question: What type of SSB mechanism can produce a graviton mass that explains the acceleration of the current Universe? Current observations suggest that dark matter contributes roughly $30\%$ of the energy budget of the Universe~\cite{2021-DiValentino.etal-Class.Quant.Grav.,Kamionkowski:2022pkx}. Although its physical nature remains unknown, it is hypothesized that cosmological phase transitions may have occurred in a dark or hidden sector through SSB, involving particles and fields that are only very weakly coupled to visible matter~\cite{Schwaller:2015tja}.  In the case of millicharged dark matter 
($g_{_A} \sim 10^{-3} e$ where $e$ is the electric charge of electron), complex scalar fields couple to $U(1)$ gauge field to generate dark photon. The experimental bound on the mass of these photons is $m_{A} < 10^{-3}~{\rm eV}$~\cite{Arias:2012az,Caputo:2021eaa}. This constraint on dark photon mass and $\lambda < 1$ leads to the following bound on graviton mass:
\[m < 10^{-28}~{\rm eV} \, .
\]
This bound is consistent with LIGO observations~\cite{abbott2021tests}, and its value is comparable to the Hubble constant $H_0 \sim 5 \times 10^{-28}~{\rm eV}$~\cite{2021-DiValentino.etal-Class.Quant.Grav.}. As mentioned earlier, a graviton mass comparable with the Hubble constant $H_{0}$ can drive the acceleration of the current Universe~\cite{deRham:2013qqa,Gershtein:1997mm}. This implies that SSB in the dark sector can yield the graviton mass needed to naturally explain the Universe's acceleration.

Lastly, to understand the contribution of the 4-rank tensor $\mathcal{O}_{\mu\nu}^{\alpha\beta}$ w.r.t the Ricci scalar and tensor during the above SSB scenario. In the case of $\Lambda$CDM model the background Einstein's equation leads to $R \sim \Lambda$. Hence, 
Ricci scalar is $\sim 10^{-64} \text{eV}^{2}$. On the other hand, the SSB in the dark matter sector (in our case) 
leads to mass squared of spin-2 field $m^{2}$ to be $10^{-56}\text{eV}^{2}$ which is $10^{8}$ order higher in magnitude. Substituting Eq.~\eqref{eq:OSSB} in Eq.~\eqref{eq. M8}, we see that effective graviton mass is $10^{-56}\text{eV}^{2}$; hence, the contribution is from the fourth-rank tensor. Note 
that the concept of mass for the spin-2 field is well-defined only if the background metric solution 
is asymptotically flat since the action for the spin-2 field asymptotically approaches the extended Fierz-Pauli action ($\alpha = \frac{1}{2}$) at spatial infinity~\cite{Mukhanov:2024mmg}.

A curious reader might wonder: Can the aforementioned mass value be generated in a more restrictive model with fewer parameters, such as the Coleman-Weinberg (CW) model~\cite{Coleman:1973jx, Jackiw:1974cv}, where $m_{\Phi} = 0$? In the CW mechanism, the model exhibits one fewer parameter and is classically scale invariant. Furthermore, the Higgs mass is directly predicted. In the rest of this section, we adopt this approach and investigate the possibility of generating the effective mass of the spin-2 field. Including the radiative quantum corrections, the one-loop effective potential is:
\begin{equation}\label{S L23}
V_{eff}(\phi_{c}) = - \frac{3g_A^{4}}{32\pi^{2}}\phi_{c}^{4}\Big[1 - 2\log\left(\frac{\phi_{c}
^{2}}{v^{2}}\right)\Big] \, ,
\end{equation}
The above potential has a maximum at $\phi_{c} = 0$, whereas a minimum at $\phi_{c} = v$.
Thus, the effective action of scalar quantum electrodynamics at one-loop level can be expressed as
\begin{equation}\label{S L25}
S_{QED}^{(1-lopp)} = \int d^{4}x \sqrt{-g} \, \Big[- \frac{1}{4}F_{\mu\nu}F^{\mu\nu} - (\mathcal{D}_{\mu}\phi)^{\dagger}(\mathcal{D}^{\mu}\phi) - V_{eff}(\phi)\Big].
\end{equation}
Considering the energy-momentum tensor and the 4-rank tensor $\mathcal{O}_{\mu\nu}^{ \ \ \rho\sigma}$ associated with the above action, we obtain the following effective action of spin-2 field at the quadratic level
\begin{equation}\label{S L29}
\begin{split}
S_{h} & = \int d^{4}x\sqrt{-\bar{g}}\Big[ 
- \frac{1}{2}\bar{\nabla}_{\lambda}h_{\mu\nu}\bar{\nabla}
 ^{\lambda}h^{\mu\nu} + \frac{1}{2}\bar{\nabla}_{\rho}h\bar{\nabla}^{\rho}h + \bar{\nabla}^{\nu}
 h^{\rho\mu}\bar{\nabla}_{\mu}h_{\rho\nu}\\ 
 & - \bar{\nabla}_{\mu}h\bar{\nabla}_{\nu}h^{\mu\nu} +  \kappa^{2}V_{eff}(\phi_{c} = v)\left(h_{\alpha\beta}h^{\alpha\beta} - \frac{1}{2}h^{2}
 \right)\Big],
\end{split}
\end{equation}
where
\begin{equation}\label{S L30}
V_{eff}(\phi_{c} = v) = - \frac{3g_A^{4}}{32\pi^{2}}v^{4} = - \frac{3}{32\pi^{2}}m_{Z}^{4}
= - \frac{M^{2}v^{2}}{8}.
\end{equation}
By repeating the above steps, we obtain the graviton mass as:
\[
m[CW] \sim \kappa \pi (m_A/g_{_A})^2  \, .
\]
Here again, for the millicharged scalar dark matter coupling to local $U(1)$ symmetry, the graviton mass turns out to be $10^{-27}~{\rm eV}$. 
This suggests that the Coleman-Weinberg mechanism in the dark sector can also yield the graviton mass required to explain the current acceleration of the Universe.

\section{Discussion} 
\label{sec:Discussion}

Our work is based on two well-established frameworks: GR and quantum field theory. Our analysis demonstrates that non-perturbative phenomena, such as spontaneous symmetry breaking (SSB), can significantly impact gravity, leading to an effective mass for spin-2 fields as a natural outcome. Although our analysis is classical, the conclusions remain robust because the energy scale of the Standard Model (SM) in particle physics is much lower than the Planck scale. As a result, quantum corrections are unlikely to alter our conclusions. While our study primarily focuses on a complex scalar field, the approach can be extended to any SSB mechanism.

Given that all matter fields carry energy and interact with gravity, it is natural to ask: Do non-perturbative phenomena like SSB affect gravity? How do masses generated through SSB influence gravity? Traditionally, gravity is considered negligible at these energy scales and is often ignored in such contexts~\cite{dawson1999introduction, hill2003strong}. While this assumption holds for perturbative effects, our results indicate that it may not apply when non-perturbative effects like SSB are considered. We have shown that SSB can induce non-trivial infrared (IR) effects on gravity. In particular, we demonstrated that spin-2 fields can acquire an \emph{effective mass} through the SSB of a global or local symmetry in a complex scalar field minimally coupled to gravity.

As mentioned earlier, it is crucial to highlight that Einstein's field equations are highly non-linear, allowing solutions only in a few highly symmetric cases. The perturbative approach discussed here is valuable because solving the Einstein equations in full generality is often not feasible. However, before performing a perturbative expansion, it is necessary to establish a background—a manifold that represents either the average universe or the background space-time. In General Relativity, this averaging process is challenging because it involves dealing with complex non-linear interactions. The impact of these non-linearities on the overall expansion is known as backreaction~\cite{Buchert:2007ik, Buchert:2011sx}. Our approach, which does not rely on any assumptions about the background metric, is particularly effective due to the non-linear nature of the potential. Even without precise knowledge of the background metric, our method effectively captures the influence of the non-linear potential on perturbations, especially in the context of SSB. As mentioned earlier, since the background metric $\bar{g}_{\mu\nu}$ is chosen to be arbitrary, it has six independent components, which also implies that the resultant metric perturbation $h_{\mu\nu}$ has 6 DoF which is mathematically consistent with our analysis. On the other hand, it is also important to note that in asymptotic spatial infinity, the DoF of metric perturbation reduces to 2 as the energy-momentum vanishes there identically. As a result, there would not be any SSB in the matter sector.

In Refs.~\cite{DeFelice:2012mx, Gumrukcuoglu:2011zh}, it is demonstrated that massive gravity permits a well-defined accelerating solution exclusively within open FLRW universes, and even in such contexts, perturbation theory reveals instabilities. In contrast to the aforementioned studies, which treat massive gravity as a modification of gravitational theory in cosmology, this work begins with GR and demonstrates that the non-zero mass of metric perturbations emerges only in the presence of SSB within the matter sector of the theory, within the framework of linearized gravity. Given that the initial gravitational theory is GR, the linearized massive gravity theory derived following the consideration of SSB explains the late-stage accelerating phase of the Universe, assuming it to be spatially flat due to the degravitation effect, as discussed in detail previously. Furthermore, it is important to emphasize that, in deriving the linearized massive gravity theory, we 
did not adopt dRGT massive gravity as the foundational theory, as is the case in Refs.~\cite{DeFelice:2012mx, Gumrukcuoglu:2011zh}.

It is important to recognize that gauge transformations are not physical symmetries, as they do not correspond to any intrinsic property of a physical system or represent physical charges. Instead, they stem from the redundancy in the formalism. On the other hand, the Higgs mechanism is a well-known method for spontaneously breaking the physical symmetry of a system. However, in the Higgs mechanism, it is not the gauge symmetry itself that is broken but rather the symmetry of the potential.
In this context, it is crucial to highlight that SSB involves concealing the symmetry through a gauge choice while the underlying gauge symmetry remains intact, as stated by Elitzur’s theorem~\cite{PhysRevD.12.3978, FROHLICH1981553, MAAS2019132, Diaz:2001yz}.
In gauge theory, physical observables are gauge-invariant and are usually computed via functional integrals over all gauge field configurations. To avoid overcounting equivalent points on a gauge orbit, a gauge-fixing procedure is employed to ensure that each gauge orbit is intersected precisely once.

The manifestation of SSB, characterized by the non-trivial vacuum expectation value of the order parameter in the matter sector, effectively represents a gauge choice. While this process grants mass to gauge bosons, the underlying gauge symmetry remains preserved via the Stueckelberg mechanism (see Appendix \ref{Appendix D}). Consequently, the effective mass acquired by spin-2 fields through the SSB mechanism does not imply the breaking of diffeomorphism invariance. 

For instance, in the case of local $U(1)$  symmetry breaking in scalar quantum electrodynamics, the non-trivial vacuum expectation value arising from SSB in the field potential leads to a non-zero mass for the $U(1)$ gauge field. This mass is physical and can be observed through the Meissner effect in superconductors (see \cite{kozhevnikov2021meissner}). While it may seem that the gauge symmetry is broken since the gauge field acquires a mass term at the action level after SSB, this is not the case, as Elitzur’s theorem ensures that the underlying gauge symmetry remains intact, as discussed earlier. In our scenario, diffeomorphism invariance is similarly preserved for any arbitrary background.

In this work, we have explicitly demonstrated that the spontaneous symmetry breaking of local and global symmetry in the matter Lagrangian density leads to a non-zero mass for spin-2 fields. Furthermore, we have established that the massive spin-2 field action corresponds to the extended FP action ($\alpha = 1/2$). Interesting $\alpha =  1/2$ leads to 6 DoF and no ghost. Among these, two are helicity-2 DoF, two are helicity-1 DoF, and two are scalar DoF. Interestingly, the scalar DoF is coupled to the trace of the energy-momentum tensor. For radiation, the trace of the energy-momentum tensor is zero; hence, the coupling will be absent. Consequently, the extra scalar degree of freedom can effectively delineate these two distinct eras through the coupling. This represents a non-trivial aspect of the massive gravity theory that is not present in the theory of massless helicity-2 fields. Such a phenomenological model was studied to explain the acceleration of the current Universe~\cite{Singh:1988pp,Sami:2002se} in terms of degravitation (see Appendix \ref{appendix: degravitation}). It will be interesting to investigate the cosmological consequence of such an analysis. 

Dark matter holds significant importance in elucidating the rotation curves of galaxies~\cite{Brownstein:2005zz, Kamada:2016euw, Persic:1995ru}.
The present work demonstrates that SSB in the dark sector naturally accounts for the small graviton mass necessary for the acceleration of the current Universe, \emph{without invoking} dark energy. For the first time, this study illuminates the intricate interplay between dark matter and mechanism of the accelerated expansion of the Universe. The implications of SSB on the IR regime of GR are a fascinating area of study that could reshape our understanding of the Universe. 

\noindent {\bf Acknowledgement:-}
The authors thank S. Uma Sankar for multiple discussions and comments on the earlier draft of the manuscript. The authors also thank I. Chakraborty, Saurya Das, and T. Parvez, for comments on the earlier draft. SM is supported by the SERB-Core Research Grant (Project SERB/CRG/2022/002348). The work is also supported by the SPARC MoE grant SPARC/2019-2020/P2926/SL.

\appendix

\section{Effective mass generation of spin-2 field in arbitrary curved spacetime}
\label{Appendix.B.new}

In this section, we give full details of the calculations presentation in the main text. The action of a complex scalar field ($\Phi$) minimally coupled to gravity in a general 4-D spacetime is:
\begin{equation}\label{eq. 6}
S = \frac{1}{2\kappa^{2}} \int \!\! d^{4}x \sqrt{-g} R 
+ S_{\Phi}(\Phi, \partial_\mu \Phi),
\end{equation}
where $\kappa^{2} = 8\pi G$. Although this is a standard calculation, for ease of verification, we compute the relevant quantities to obtain the linearised theory of gravity within the framework of Einstein's general relativity in this supplemental material. In order to do that, we consider the following decomposition of the metric
\begin{equation}\label{app.metric decomposition 1}
g_{\mu\nu} = \bar{g}_{\mu\nu} + \varepsilon h_{\mu\nu},
\end{equation}
where $\bar{g}_{\mu\nu}$ is the metric describing the background geometry \textit{w.r.t}
which we linearised the theory of gravity, $h_{\mu\nu}$ is the metric perturbation and
$\varepsilon (= 2\kappa)$ carries the order of the metric perturbation. This leads to the following
expression of the inverse metric:
\begin{equation}\label{inverse metric 1}
g^{\mu\nu} = \bar{g}^{\mu\nu} - \varepsilon h^{\mu\nu} + \varepsilon^{2}h^{\mu\rho}
h_{ \ \rho}^{\nu} + \mathcal{O}(\varepsilon^{3}),
\end{equation}
where $\bar{g}^{\mu\nu}$ is the inverse metric of the background spacetime, and $h^{\mu\nu}$ is a rank-2 contravariant tensor defined by contracting the metric perturbation \textit{w.r.t} the inverse metric $\bar{g}^{\mu\nu}$. Given the above set of relations, we are now able to compute the Christoffel symbol defined in the following manner
\begin{equation}\label{Christoffel symbol 1}
\Gamma_{ \ \mu\nu}^{\rho} = \frac{1}{2}g^{\rho\lambda}(\partial_{\mu}g_{\lambda\nu} + 
\partial_{\nu}g_{\mu\lambda} - \partial_{\lambda}g_{\mu\nu}),
\end{equation} 
which reduces to the following form after using the relations \eqref{app.metric decomposition 1} and \eqref{inverse metric 1}
\begin{equation}\label{S.30}
\begin{split}
\Gamma_{ \ \mu\nu}^{\rho} & = \bar{\Gamma}_{ \ \mu\nu}^{\rho} + \varepsilon\Big[\frac{1}{2}
\bar{g}^{\rho\lambda}(\partial_{\mu}h_{\lambda\nu} + \partial_{\nu}h_{\mu\lambda} - 
\partial_{\lambda}h_{\mu\nu})\\
 & - \bar{g}^{\rho\alpha}h_{\alpha\beta}\bar{\Gamma}_{ \ \mu\nu}^{\beta}\Big]\\
 &  - \varepsilon^{2}\Big[\frac{1}{2}\bar{g}^{\beta\lambda}h_{ \ \beta}^{\rho}(\partial_{\mu}
 h_{\lambda\nu} + \partial_{\nu}h_{\mu\lambda} - \partial_{\lambda}h_{\mu\nu})\\
 & - \bar{g}^{\alpha\beta}h_{ \ \beta}^{\rho}h_{\alpha\sigma}\bar{\Gamma}_{ \ \mu\nu}^{\sigma}\Big], 
\end{split}
\end{equation}
where
\begin{equation}\label{S.31}
\bar{\Gamma}_{ \ \mu\nu}^{\rho} = \frac{1}{2}\bar{g}^{\rho\lambda}(\partial_{\mu}\bar{g}
_{\lambda\nu} + \partial_{\nu}\bar{g}_{\mu\lambda} - \partial_{\lambda}\bar{g}_{\mu\nu}).
\end{equation}
After a change in the indices of the following form
\begin{equation}\label{S.32}
\begin{split}
\bar{g}^{\rho\alpha}h_{\alpha\beta}\bar{\Gamma}_{ \ \mu\nu}^{\beta} & \rightarrow 
\bar{g}^{\rho\lambda}h_{\lambda\beta}\bar{\Gamma}_{ \ \mu\nu}^{\beta}\\
\bar{g}^{\alpha\beta}h_{ \ \beta}^{\rho}h_{\alpha\sigma}\bar{\Gamma}_{ \ \mu\nu}^{\sigma}
& \rightarrow \bar{g}^{\lambda\beta}h_{ \ \beta}^{\rho}h_{\lambda\sigma}\bar{\Gamma}_{ \ 
\mu\nu}^{\sigma},
\end{split}
\end{equation}
we obtain the following expression of Christoffel symbols
\begin{equation}\label{S.33}
\begin{split}
\Gamma_{ \ \mu\nu}^{\rho} & = \bar{\Gamma}_{ \ \mu\nu}^{\rho} + \frac{\varepsilon}{2}
\bar{g}^{\rho\lambda}[\partial_{\mu}h_{\lambda\nu} + \partial_{\nu}h_{\mu\lambda} - 
\partial_{\lambda}h_{\mu\nu} - 2h_{\lambda\beta}\bar{\Gamma}_{ \ \mu\nu}^{\beta}]\\
 & - \frac{\varepsilon^{2}}{2}\bar{g}^{\beta\lambda}h_{ \ \beta}^{\rho}[\partial_{\mu}
 h_{\lambda\nu} + \partial_{\nu}h_{\mu\lambda} - \partial_{\lambda}h_{\mu\nu} - 
 2h_{\lambda\sigma}\bar{\Gamma}_{ \ \mu\nu}^{\sigma}].
\end{split}
\end{equation}
Now adding and subtracting the following terms at different order
\begin{equation}\label{S.34}
\begin{split}
\mathcal{O}(\varepsilon): & h_{\mu\beta}\bar{\Gamma}_{ \ \nu\lambda}^{\beta} + h_{\nu\beta}
\bar{\Gamma}_{ \ \mu\lambda}^{\beta}\\
\mathcal{O}(\varepsilon^{2}): & h_{\mu\sigma}\bar{\Gamma}_{ \ \nu\lambda}^{\sigma} + 
h_{\nu\sigma}\bar{\Gamma}_{ \ \mu\lambda}^{\sigma}, 
\end{split}
\end{equation}
and then regrouping the terms gives
\begin{equation}\label{S.35}
\begin{split}
\Gamma_{ \ \mu\nu}^{\rho} & = \bar{\Gamma}_{ \ \mu\nu}^{\rho} + \frac{\varepsilon}{2}
\bar{g}^{\rho\lambda}(\bar{\nabla}_{\mu}h_{\lambda\nu} + \bar{\nabla}_{\nu}h_{\mu\lambda}
- \bar{\nabla}_{\lambda}h_{\mu\nu})\\
 & - \frac{\varepsilon^{2}}{2}h_{ \ \beta}^{\rho}\bar{g}^{\beta\lambda}(\bar{\nabla}_{\mu}
 h_{\lambda\nu} + \bar{\nabla}_{\nu}h_{\mu\lambda} - \bar{\nabla}_{\lambda}h_{\mu\nu}),
\end{split}
\end{equation}
where we recognize the following
\begin{equation}\label{S.36}
\bar{\nabla}_{\lambda}h_{\mu\nu} = \partial_{\lambda}h_{\mu\nu} - h_{\nu\sigma}\bar{\Gamma}
_{ \ \mu\lambda}^{\sigma} - h_{\mu\sigma}\bar{\Gamma}_{ \ \nu\lambda}^{\sigma}.
\end{equation}
Therefore, defining the linearised Christoffel symbol as
\begin{equation}\label{S.37}
(\Gamma_{ \ \mu\nu}^{\rho})_{L} = \frac{1}{2}\bar{g}^{\rho\lambda}(\bar{\nabla}_{\mu}
h_{\lambda\nu} + \bar{\nabla}_{\nu}h_{\mu\lambda} - \bar{\nabla}_{\lambda}h_{\mu\nu}),
\end{equation}
we may now write
\begin{equation}\label{S.38}
\Gamma_{ \ \mu\nu}^{\rho} = \bar{\Gamma}_{ \ \mu\nu}^{\rho} + \varepsilon(\Gamma_{ \ \mu\nu}
^{\rho})_{L} - \varepsilon^{2}h_{ \ \beta}^{\rho}(\Gamma_{ \ \mu\nu}^{\beta})_{L}.
\end{equation}
The Riemann curvature tensor is defined as
\begin{equation}\label{S.39}
R_{ \ \nu\rho\sigma}^{\mu} = \partial_{\rho}\Gamma_{ \ \sigma\nu}^{\mu} + \Gamma_{ \ \rho
\lambda}^{\mu}\Gamma_{ \ \sigma\nu}^{\lambda} - \partial_{\sigma}\Gamma_{ \ \rho\nu}^{\mu}
- \Gamma_{ \ \sigma\lambda}^{\mu}\Gamma_{ \ \rho\nu}^{\lambda}.
\end{equation}
We start by defining the following quantity
\begin{equation}\label{defintion 1}
\delta\Gamma_{ \ \mu\nu}^{\rho} \equiv \Gamma_{ \ \mu\nu}^{\rho} - \bar{\Gamma}_{ \ \mu\nu}
^{\rho} = \varepsilon(\Gamma_{ \ \mu\nu}^{\rho})_{L} - \varepsilon^{2}h_{ \ \beta}^{\rho}
(\Gamma_{ \ \mu\nu}^{\beta})_{L},
\end{equation}
in terms of which we may express the Riemann curvature tensor as
\begin{equation}\label{Riemann tensor decomposition 1}
\begin{split}
R_{ \ \nu\rho\sigma}^{\mu} & = \bar{R}_{ \ \nu\rho\sigma}^{\mu} + (\partial_{\rho}\delta
\Gamma_{ \ \sigma\nu}^{\mu} + \bar{\Gamma}_{ \ \rho\lambda}^{\mu}\delta\Gamma_{ \ \sigma
\nu}^{\lambda} - \bar{\Gamma}_{ \ \rho\nu}^{\lambda}\delta\Gamma_{ \ \sigma\lambda}^{\mu})\\
 & - (\partial_{\sigma}\delta\Gamma_{ \ \rho\nu}^{\mu} + \bar{\Gamma}_{ \ \sigma\lambda}
 ^{\mu}\delta\Gamma_{ \ \rho\nu}^{\lambda} - \bar{\Gamma}_{ \ \sigma\nu}^{\lambda}\delta
 \Gamma_{ \ \rho\lambda}^{\mu})\\
 & + \delta\Gamma_{ \ \rho\lambda}^{\mu}\delta\Gamma_{ \ \sigma\nu}^{\lambda} - \delta\Gamma
_{ \ \sigma\lambda}^{\mu}\delta\Gamma_{ \ \rho\nu}^{\lambda},
\end{split}
\end{equation}
where
\begin{equation}\label{S.40}
\bar{R}_{ \ \nu\rho\sigma}^{\mu} = \partial_{\rho}\bar{\Gamma}_{ \ \sigma\nu}^{\mu} + 
\bar{\Gamma}_{ \ \rho\lambda}^{\mu}\bar{\Gamma}_{ \ \sigma\nu}^{\lambda} - \partial_{\sigma}
\bar{\Gamma}_{ \ \rho\nu}^{\mu} - \bar{\Gamma}_{ \ \sigma\lambda}^{\mu}\bar{\Gamma}_{ \ \rho
\nu}^{\lambda}.
\end{equation}
Now adding and subtracting the term $\bar{\Gamma}_{ \ \rho\sigma}^{\lambda}\delta\Gamma_{ \ 
\lambda\nu}^{\mu}$ from the equation \eqref{Riemann tensor decomposition 1}, we obtain the 
following relation
\begin{equation}\label{S.41}
R_{ \ \nu\rho\sigma}^{\mu} = \bar{R}_{ \ \nu\rho\sigma}^{\mu} + \bar{\nabla}_{\rho}\delta
\Gamma_{ \ \sigma\nu}^{\mu} - \bar{\nabla}_{\sigma}\delta\Gamma_{ \ \rho\nu}^{\mu} + \delta
\Gamma_{ \ \rho\lambda}^{\mu}\delta\Gamma_{ \ \sigma\nu}^{\lambda} - \delta\Gamma_{ \ \sigma
\lambda}^{\mu}\delta\Gamma_{ \ \rho\nu}^{\lambda}. 
\end{equation}
Now using the relation \eqref{defintion 1}, we now express the above relation as follows
\begin{equation}\label{S.42}
\begin{split}
R_{ \ \nu\rho\sigma}^{\mu} & = \bar{R}_{ \ \nu\rho\sigma}^{\mu} + \varepsilon[\bar{\nabla}
_{\rho}(\Gamma_{ \ \sigma\nu}^{\mu})_{L} - \bar{\nabla}_{\sigma}(\bar{\Gamma}_{ \ \rho\nu}
^{\mu})_{L}]\\
 & - \varepsilon^{2}\{\bar{\nabla}_{\rho}[h_{ \ \beta}^{\mu}(\Gamma_{ \ \sigma\nu}^{\beta})
_{L}] - \bar{\nabla}_{\sigma}[h_{ \ \beta}^{\mu}(\Gamma_{ \ \rho\nu}^{\beta})_{L}]\\
 & - (\Gamma_{ \ \rho\lambda}^{\mu})_{L}(\Gamma_{ \ \sigma\nu}^{\lambda})_{L} + (\Gamma_{ \ 
 \sigma\lambda}^{\mu})_{L}(\Gamma_{ \ \rho\nu}^{\lambda})_{L}\} + \mathcal{O}(\varepsilon^{3}).
\end{split}
\end{equation}
The above expression can further be reduced in the following form
\begin{equation}\label{S.43}
\begin{split}
R_{ \ \nu\rho\sigma}^{\mu} & = \bar{R}_{ \ \nu\rho\sigma}^{\mu} + \varepsilon\{\bar{\nabla}
_{\rho}(\Gamma_{ \ \sigma\nu}^{\mu})_{L} - \bar{\nabla}_{\sigma}(\Gamma_{ \ \rho\nu}^{\mu})
_{L}\}\\
 & - \varepsilon^{2}\{h_{ \ \beta}^{\mu}[\bar{\nabla}_{\rho}(\Gamma_{ \ \sigma\nu}^{\beta})
_{L} - \bar{\nabla}_{\sigma}(\Gamma_{ \ \rho\nu}^{\beta})_{L}]\\
 & + (\Gamma_{ \ \sigma\nu}^{\beta})_{L}\bar{\nabla}_{\rho}h_{ \ \beta}^{\mu} - (\Gamma_{ \ 
\rho\nu}^{\beta})_{L}\bar{\nabla}_{\sigma}h_{ \ \beta}^{\mu}\\
 & - (\Gamma_{ \ \rho\beta}^{\mu})_{L}(\Gamma_{ \ \sigma\nu}^{\beta})_{L} + (\Gamma_{ \ 
\sigma\beta}^{\mu})_{L}(\Gamma_{ \ \rho\nu}^{\beta})_{L}\}.
\end{split}
\end{equation}
Defining the following quantity
\begin{equation}\label{S.44}
(R_{ \ \nu\rho\sigma}^{\mu})_{L} = \bar{\nabla}_{\rho}(\Gamma_{ \ \sigma\nu}^{\mu})_{L} - 
\bar{\nabla}_{\sigma}(\Gamma_{ \ \rho\nu}^{\mu})_{L},
\end{equation}
we further reduce the expression of Riemann curvature tensor as
\begin{equation}\label{S.45}
\begin{split}
R_{ \ \nu\rho\sigma}^{\mu} & = \bar{R}_{ \ \nu\rho\sigma}^{\mu} + \varepsilon(R_{ \ \nu\rho
\sigma}^{\mu})_{L} - \varepsilon^{2}\{h_{ \ \beta}^{\mu}(R_{ \ \nu\rho\sigma}^{\beta})_{L}\\
 & + (\Gamma_{ \ \sigma\nu}^{\beta})_{L}[\bar{\nabla}_{\rho}h_{ \ \beta}^{\mu} - (\Gamma_{ \ 
 \rho\beta}^{\mu})_{L}]\\
 & - (\Gamma_{ \ \rho\nu}^{\beta})_{L}[\bar{\nabla}_{\sigma}h_{ \ \beta}^{\mu} - (\Gamma_{ \ 
\sigma\beta}^{\mu})_{L}]\}.
\end{split}
\end{equation}
The term $* \equiv (\Gamma_{ \ \sigma\nu}^{\beta})_{L}[\bar{\nabla}_{\rho}h_{ \ \beta}^{\mu} 
- (\Gamma_{ \ \rho\beta}^{\mu})_{L}]$ can be expressed as
\begin{equation}\label{S.46}
\begin{split}
* & = (\Gamma_{ \ \sigma\nu}^{\beta})_{L}\Big[\bar{\nabla}_{\rho}(\bar{g}^{\mu\alpha}h_{\alpha
\beta}) - \frac{1}{2}\bar{g}^{\mu\alpha}(\bar{\nabla}_{\rho}h_{\alpha\beta} + \bar{\nabla}_{\beta}
h_{\rho\alpha}\\
 & - \bar{\nabla}_{\alpha}h_{\rho\beta})\Big]\\
 & = (\Gamma_{ \ \sigma\nu}^{\beta})_{L}\Big[\frac{1}{2}\bar{g}^{\mu\alpha}(\bar{\nabla}_{\rho}
h_{\alpha\beta} + \bar{\nabla}_{\alpha}h_{\rho\beta} - \bar{\nabla}_{\beta}h_{\rho\alpha})\Big]\\
 & = (\Gamma_{ \ \sigma\nu}^{\beta})_{L}\Big[\frac{1}{2}\bar{g}^{\mu\alpha}\delta_{\beta}^{ \ 
\lambda}(\bar{\nabla}_{\rho}h_{\alpha\lambda} + \bar{\nabla}_{\alpha}h_{\rho\lambda} - \bar{\nabla}
_{\lambda}h_{\rho\alpha})\Big]\\
 & = (\Gamma_{ \ \sigma\nu}^{\beta})_{L}\Big[\frac{1}{2}\bar{g}^{\mu\alpha}\bar{g}_{\beta\gamma}
 \bar{g}^{\gamma\lambda}(\bar{\nabla}_{\rho}h_{\alpha\lambda} + \bar{\nabla}_{\alpha}h_{\rho\lambda} 
 - \bar{\nabla}_{\lambda}h_{\rho\alpha})\Big]\\
 & = \bar{g}^{\mu\alpha}\bar{g}_{\beta\gamma}(\Gamma_{ \ \sigma\nu}^{\beta})_{L}(\Gamma_{ \ \rho
 \alpha}^{\gamma})_{L}.
\end{split}
\end{equation}
In the similar manner, we may express the following
\begin{equation}\label{S.47}
(\Gamma_{ \ \rho\nu}^{\beta})_{L}[\bar{\nabla}_{\sigma}h_{ \ \beta}^{\mu} - (\Gamma_{ \ \sigma
\beta}^{\mu})_{L}] = \bar{g}^{\mu\alpha}\bar{g}_{\beta\gamma}(\Gamma_{ \ \rho\nu}^{\beta})_{L}
(\Gamma_{ \ \sigma\alpha}^{\gamma})_{L}.
\end{equation}
Using the above results, the final form of the Riemann tensor perturbed up to second order in 
metric perturbation is given as
\begin{equation}\label{S.48}
\begin{split}
R_{ \ \nu\rho\sigma}^{\mu} & = \bar{R}_{ \ \nu\rho\sigma}^{\mu} + \varepsilon(R_{ \ \nu\rho\sigma}^{\mu})_{L} - \varepsilon^{2}\{h_{ \ \beta}^{\mu}(R_{ \ \nu\rho\sigma}^{\beta})_{L}\\
 & + \bar{g}^{\mu\alpha}\bar{g}_{\beta\gamma}[(\Gamma_{ \ \sigma\nu}^{\beta})_{L}(\Gamma_{ \ \rho
\alpha}^{\gamma})_{L} - (\Gamma_{ \ \rho\nu}^{\beta})_{L}(\Gamma_{ \ \sigma\alpha}^{\gamma})_{L}]\}.
\end{split}
\end{equation}
As a result, the Ricci tensor can be expressed as
\begin{equation}\label{S.49}
\begin{split}
R_{\nu\sigma} & = \bar{R}_{\nu\sigma} + \varepsilon(R_{\nu\sigma})_{L} - \varepsilon^{2}
\{h_{ \ \beta}^{\mu}(R_{ \ \nu\mu\sigma}^{\beta})_{L}\\
 & + \bar{g}^{\mu\alpha}\bar{g}_{\beta\gamma}[(\Gamma_{ \ \sigma\nu}^{\beta})_{L}(\Gamma_{ \ \mu
\alpha}^{\gamma})_{L} - (\Gamma_{ \ \mu\nu}^{\beta})_{L}(\Gamma_{ \ \sigma\alpha}^{\gamma})_{L}]\}.
\end{split}
\end{equation}
From the definition of linearised Riemann tensor, we have the following expression
\begin{equation}\label{S.50}
\begin{split}
(R_{ \ \nu\rho\sigma}^{\mu})_{L} & = \frac{1}{2}[\bar{\nabla}_{\rho}\bar{\nabla}_{\sigma}h_{ \ 
\nu}^{\mu} + \bar{\nabla}_{\rho}\bar{\nabla}_{\nu}h_{ \ \sigma}^{\mu} - \bar{\nabla}_{\rho}
\bar{\nabla}^{\mu}h_{\sigma\nu}\\
 & - \bar{\nabla}_{\sigma}\bar{\nabla}_{\rho}h_{ \ \nu}^{\mu} - \bar{\nabla}_{\sigma}\bar{\nabla}
_{\nu}h_{ \ \rho}^{\mu} + \bar{\nabla}_{\sigma}\bar{\nabla}^{\mu}h_{\rho\nu}],
\end{split}
\end{equation}
which leads to the following expression of $(R_{\nu\sigma})_{L}$
\begin{equation}\label{S.51}
\begin{split}
(R_{\nu\sigma})_{L} & = (R_{ \ \nu\mu\sigma}^{\mu})_{L}\\
 & = \frac{1}{2}(\bar{\nabla}_{\mu}\bar{\nabla}_{\sigma}h_{ \ \nu}^{\mu} + \bar{\nabla}_{\mu}
 \bar{\nabla}_{\nu}h_{ \ \sigma}^{\mu} - \bar{\Box}h_{\sigma\nu} - \bar{\nabla}_{\sigma}\bar{
 \nabla}_{\nu}h).
\end{split}
\end{equation}
Therefore, the Ricci scalar can now be expressed as
\begin{equation}\label{S.52}
\begin{split}
R & = (\bar{g}^{\nu\sigma} - \varepsilon h^{\nu\sigma} + \varepsilon^{2}h^{\nu\lambda}h_{ \ 
\lambda}^{\sigma})[\bar{R}_{\nu\sigma} + \varepsilon(R_{\nu\sigma})_{L}\\
 & - \varepsilon^{2}\{h_{ \ \beta}^{\mu}(R_{ \ \nu\mu\sigma}^{\beta})_{L} + \bar{g}^{\mu\alpha}
 \bar{g}_{\beta\gamma}[(\Gamma_{ \ \sigma\nu}^{\beta})_{L}(\Gamma_{ \ \mu\alpha}^{\gamma})_{L}\\
 & - (\Gamma_{ \ \mu\nu}^{\beta})_{L}(\Gamma_{ \ \sigma\alpha}^{\gamma})_{L}]\}]\\
 & = \bar{g}^{\nu\sigma}\bar{R}_{\nu\sigma} + \varepsilon[\bar{g}^{\nu\sigma}(R_{\nu\sigma})_{L}
 - h^{\nu\sigma}\bar{R}_{\nu\sigma}]\\
 & - \varepsilon^{2}[\bar{g}^{\nu\sigma}\{h_{ \ \beta}^{\mu}(R_{ \ \nu\mu\sigma}^{\beta})_{L} + 
 \bar{g}^{\mu\alpha}\bar{g}_{\beta\gamma}[(\Gamma_{ \ \sigma\nu}^{\beta})_{L}(\Gamma_{ \ \mu\alpha}
^{\gamma})_{L}\\
 & - (\Gamma_{ \ \mu\nu}^{\beta})_{L}(\Gamma_{ \ \sigma\alpha}^{\gamma})_{L}]\} + h^{\nu\sigma}
 (R_{\nu\sigma})_{L} - h^{\nu\lambda}h_{ \ \lambda}^{\sigma}\bar{R}_{\nu\sigma}] + \mathcal{O}
(\varepsilon^{3}).
\end{split}
\end{equation}
Using the following definitions
\begin{equation}\label{S.53}
\bar{R} = \bar{g}^{\nu\sigma}\bar{R}_{\nu\sigma}, \ R_{L} = \bar{g}^{\nu\sigma}(R_{\nu\sigma})_{L}
- h^{\nu\sigma}\bar{R}_{\nu\sigma},
\end{equation}
and neglecting the higher order contributions, we obtain the following form of Ricci scalar
\begin{equation}\label{S.54}
\begin{split}
R & = \bar{R} + \varepsilon R_{L} - \varepsilon^{2}\{\bar{g}^{\nu\sigma}h_{ \ \beta}^{\mu}
(R_{ \ \nu\mu\sigma}^{\beta})_{L} + h^{\nu\sigma}(R_{\nu\sigma})_{L}\\
 & - h^{\nu\lambda}h_{ \ \lambda}^{\sigma}\bar{R}_{\nu\sigma} + \bar{g}^{\nu\sigma}\bar{g}
^{\mu\alpha}\bar{g}_{\beta\gamma}[(\Gamma_{ \ \sigma\nu}^{\beta})_{L}(\Gamma_{ \ \mu\alpha}
^{\gamma})_{L} - (\Gamma_{ \ \mu\nu}^{\beta})_{L}(\Gamma_{ \ \sigma\alpha}^{\gamma})_{L}]\}.
\end{split}
\end{equation}
We may write the above expression of Ricci scalar as
\begin{equation}\label{S.55}
\begin{split}
R & = \bar{R} + \varepsilon[\bar{g}^{\nu\sigma}(R_{\nu\sigma})_{L} - h^{\nu\sigma}
\bar{R}_{\nu\sigma}]  - \varepsilon^{2}\{\bar{g}^{\nu\sigma}h_{ \ \beta}^{\mu}
(R_{ \ \nu\mu\sigma}^{\beta})_{L}\\
 & + h^{\nu\sigma}(R_{\nu\sigma})_{L} - h^{\nu\lambda}h_{ \ \lambda}^{\sigma}\bar{R}_{\nu\sigma} + \bar{g}^{\nu\sigma}\bar{g}^{\mu\alpha}\bar{g}_{\beta\gamma}[(\Gamma_{ \ \sigma\nu}^{\beta})_{L}(\Gamma_{ \ \mu\alpha}^{\gamma})_{L}\\
 & - (\Gamma_{ \ \mu\nu}^{\beta})_{L}(\Gamma_{ \ \sigma\alpha}^{\gamma})_{L}]\}\\
 & = \bar{R} + \varepsilon\Big[\frac{1}{2}\bar{g}^{\nu\sigma}(\bar{\nabla}_{\mu}\bar{\nabla}
 _{\sigma}h_{ \ \nu}^{\mu} + \bar{\nabla}_{\mu}\bar{\nabla}_{\nu}h_{ \ \sigma}^{\mu} - 
 \bar{\Box}h_{\sigma\nu} - \bar{\nabla}_{\sigma}\bar{\nabla}_{\nu}h)\\
 & - h^{\nu\sigma}\bar{R}_{\nu\sigma}\Big] - \varepsilon^{2}\Bigg[\frac{1}{2}\bar{g}^{\nu\sigma}
 h_{ \ \mu}^{\rho}[\bar{\nabla}_{\rho}\bar{\nabla}_{\sigma}h_{ \ \nu}^{\mu} + \bar{\nabla}_{\rho}\bar{\nabla}_{\nu}h_{ \ \sigma}^{\mu} - \bar{\nabla}_{\rho}\bar{\nabla}^{\mu}h_{\sigma\nu}\\
 & - \bar{\nabla}_{\sigma}\bar{\nabla}_{\rho}
 h_{ \ \nu}^{\mu} - \bar{\nabla}_{\sigma}\bar{\nabla}_{\nu}h_{ \ \rho}^{\mu} + \bar{\nabla}_{\sigma}\bar{\nabla}^{\mu}h_{\rho\nu}]\\
 & + \frac{1}{2}h^{\nu\sigma}(\bar{\nabla}_{\mu}\bar{\nabla}_{\sigma}h_{ \ \nu}^{\mu} + \bar{\nabla}
 _{\mu}\bar{\nabla}_{\nu}h_{ \ \sigma}^{\mu} - \bar{\Box}h_{\sigma\nu} - \bar{\nabla}_{\sigma}\bar{
 \nabla}_{\nu}h)\\
 & - h^{\nu\lambda}h_{ \ \lambda}^{\sigma}\bar{R}_{\nu\sigma} + \frac{1}{4}\Big[4\bar{\nabla}
 _{\sigma}h_{\rho}^{ \ \sigma}\bar{\nabla}_{\mu}h^{\rho\mu} + 2\bar{\nabla}^{\mu}h^{\lambda\sigma}
 \bar{\nabla}_{\lambda}h_{\sigma\mu}\\
 & - 4\bar{\nabla}_{\sigma}h^{\lambda\sigma}\bar{\nabla}_{\lambda}h
 + \bar{\nabla}_{\rho}h\bar{\nabla}^{\rho}h - 3\bar{\nabla}_{\rho}h_{\mu\nu}\bar{\nabla}^{\rho}
 h^{\mu\nu}\Big]\Bigg].
\end{split}
\end{equation}
At this point, we want to mention the following relation
\begin{equation}\label{S.56}
\sqrt{-g} = \sqrt{- \bar{g}}\Big[1 + \frac{1}{2}\varepsilon h - \frac{\varepsilon^{2}}{4}
h_{\alpha\beta}h^{\alpha\beta} + \frac{\varepsilon^{2}}{8}h^{2}\Big],
\end{equation}
and since we are interested in quadratic order in metric perturbation, we can consider integration by 
parts in the quadratic order expression of Ricci scalar mentioned above. As a result, we may write 
\begin{equation}\label{S.57}
\begin{split}
\sqrt{-g}R & = \sqrt{-\bar{g}}\Bigg[\bar{R} + \varepsilon\left(\frac{1}{2}h\bar{R} - h^{\nu\sigma}\bar{R}_{\nu\sigma}\right) + \varepsilon\underbrace{\Big[\bar{\nabla}_{\mu}\bar{\nabla}_{\nu}h^{\mu\nu} - \bar{\Box}h\Big]}_{\text{surface terms}}\\
 & + \frac{\varepsilon^{2}}{2}h\Big[\bar{\nabla}_{\mu}\bar{\nabla}_{\nu}h^{\mu\nu} - \bar{\Box}h\Big]\\
 & - \frac{\varepsilon^{2}}{4}\left(h_{\alpha\beta}h^{\alpha\beta} - \frac{1}{2}h^{2}\right)\bar{R} + 
 \varepsilon^{2}\bar{R}_{\nu\sigma}\left(h^{\nu\lambda}h_{ \ \lambda}^{\sigma} - \frac{1}{2}h
 h^{\nu\sigma}\right)\\
 & - \varepsilon^{2}\left(\frac{1}{4}\bar{\nabla}_{\lambda}h_{\mu\nu}\bar{\nabla}^{\lambda}h^{\mu\nu} 
 + \frac{1}{4}\bar{\nabla}_{\rho}h\bar{\nabla}^{\rho}h - \frac{1}{2}\bar{\nabla}^{\nu}h^{\rho\mu}
 \bar{\nabla}_{\mu}h_{\rho\nu}\right)\Bigg]\\
 & = \sqrt{-\bar{g}}\Bigg[\bar{R} - \varepsilon h^{\nu\sigma}\bar{G}_{\nu\sigma} - \frac{\varepsilon^{2}}{4}\left(h_{\alpha\beta}h^{\alpha\beta} - \frac{1}{2}h^{2}\right)\bar{R}\\
 & + \varepsilon^{2}\bar{R}_{\nu\sigma}\left(h^{\nu\lambda}h_{ \ \lambda}^{\sigma} - \frac{1}{2}hh^{\nu\sigma}\right) - \varepsilon^{2}\Big[\frac{1}{4}\bar{\nabla}_{\lambda}h_{\mu\nu}\bar{\nabla}^{\lambda}h^{\mu\nu}\\
 & - \frac{1}{4}\bar{\nabla}_{\rho}h\bar{\nabla}^{\rho}h - \frac{1}{2}\bar{\nabla}^{\nu}h^{\rho\mu}\bar{\nabla}_{\mu}h_{\rho\nu} + \frac{1}{2}\bar{\nabla}_{\mu}h\bar{\nabla}_{\nu}h^{\mu\nu}\Big]\Bigg].
\end{split}
\end{equation}
Now considering $\varepsilon = \kappa$ and rescaling the action of spin-2 field by a factor 
of 4 (this is equivalent to taking $\varepsilon = 2\kappa$ and considering the vacuum Einstein field equations), the linearised Einstein-Hilbert action (without matter) in arbitrary curved spacetime can be expressed as
\begin{equation}\label{S.58}
\begin{split}
S_{EH} & = \frac{1}{2\kappa^{2}}\int d^{4}x \sqrt{-g}R = \frac{1}{2\kappa^{2}}\int d^{4}x 
\sqrt{-\bar{g}}\bar{R}\\
 & + \int d^{4}x\sqrt{-\bar{g}}\Big[ - \frac{1}{2}\bar{\nabla}_{\lambda}h_{\mu\nu}\bar{\nabla}
 ^{\lambda}h^{\mu\nu} + \frac{1}{2}\bar{\nabla}_{\rho}h\bar{\nabla}^{\rho}h\\
 & + \bar{\nabla}^{\nu}
 h^{\rho\mu}\bar{\nabla}_{\mu}h_{\rho\nu} - \bar{\nabla}_{\mu}h\bar{\nabla}_{\nu}h^{\mu\nu} 
 - \frac{1}{2}\bar{R}\left(h_{\alpha\beta}h^{\alpha\beta} - \frac{1}{2}h^{2}\right)\\
 & + 2\bar{R}_{\nu\sigma}\left(h^{\nu\lambda}h_{ \ \lambda}^{\sigma} - \frac{1}{2}h
 h^{\nu\sigma}\right)\Big]. 
\end{split}
\end{equation}

On the other hand, if we consider matter to be there, then we may write the following
\begin{equation}\label{S.60}
\begin{split}
S_{M}[\Phi, \Phi^{*}, g_{\mu\nu}] & = S_{M}[\Phi, \Phi^{*}, \bar{g}_{\mu\nu}] + 
\kappa\int\sqrt{-\bar{g}}h_{\mu\nu}T^{\mu\nu}\\
 & - \kappa^{2}\int d^{4}x\sqrt{-\bar{g}}h^{\mu\nu}h_{\rho\sigma}\mathcal{O}_{\mu\nu}^{ \ \ \rho\sigma} + \ldots.
\end{split}
\end{equation}
We may note that the linear part cancels with the second term in the first line of the 
equation (\ref{S.57}) since this is nothing but the Einstein's equation in the presence
of matter source. Moreover, it is mathematically consistent since if we expand the total
matter and Einstein-Hilbert action around a solution of classical equation of motion, the linear terms drop out and leading order terms are quadratic order in the perturbations. 

If we rescale again the spin-2 field contribution by a factor of $4$ coming from the 
matter sector, we obtain the following total action in the spin-2 field in curved spacetime
\begin{equation}\label{S.61}
\begin{split}
S_{h} & = \int d^{4}x\sqrt{-\bar{g}}\Big[ - \frac{1}{2}\bar{\nabla}_{\lambda}h_{\mu\nu}\bar{\nabla}
 ^{\lambda}h^{\mu\nu} + \frac{1}{2}\bar{\nabla}_{\rho}h\bar{\nabla}^{\rho}h\\
 & + \bar{\nabla}^{\nu}
 h^{\rho\mu}\bar{\nabla}_{\mu}h_{\rho\nu} - \bar{\nabla}_{\mu}h\bar{\nabla}_{\nu}h^{\mu\nu} 
 - \frac{1}{2}\bar{R}\left(h_{\alpha\beta}h^{\alpha\beta} - \frac{1}{2}h^{2}\right)\\
 & + 2\bar{R}_{\nu\sigma}\left(h^{\nu\lambda}h_{ \ \lambda}^{\sigma} - \frac{1}{2}h
 h^{\nu\sigma}\right)\Big] - \kappa^{2}\int d^{4}x\sqrt{-\bar{g}}h^{\mu\nu}h_{\rho\sigma}
 \mathcal{O}_{\mu\nu}^{ \ \ \rho\sigma},
\end{split}
\end{equation}
where
\begin{equation}\label{S.62}
\mathcal{O}_{\mu\nu}^{ \ \ \rho\sigma} = \frac{\delta T_{\mu\nu}}{\delta\bar{g}_{\rho\sigma}}
- \frac{1}{2}\bar{g}^{\rho\sigma}T_{\mu\nu}.
\end{equation}
As we discussed in the main text for a complex scalar field theory with local $U(1)$ gauge 
invariance
\begin{equation}\label{S.63}
\begin{split}
T_{\mu\nu} & = \mathcal{D}_{\mu}\Phi\mathcal{D}_{\nu}\Phi^{*} + \mathcal{D}_{\nu}\Phi\mathcal{D}
_{\mu}\Phi^{*} + F_{\mu\alpha}F_{\nu\beta}\bar{g}^{\alpha\beta}\\
 & - \frac{1}{4}\bar{g}_{\mu\nu}F_{\rho\sigma}F_{\alpha\beta}\bar{g}^{\rho\alpha}\bar{g}^{\beta\sigma}
 - \bar{g}_{\mu\nu}[\bar{g}^{\rho\sigma}\mathcal{D}_{\rho}\Phi\mathcal{D}_{\sigma}\Phi^{*} + V(|\Phi|)],
\end{split}
\end{equation}
which also leads to the following expression
\begin{equation}\label{S.64}
\begin{split}
\mathcal{O}_{\mu\nu}^{ \ \ \rho\sigma} & = - \frac{1}{2}F_{\mu\alpha}F_{\nu\beta}(\bar{g}
^{\alpha\rho}\bar{g}^{\beta\sigma} + \bar{g}^{\alpha\sigma}\bar{g}^{\beta\rho}) - \frac{1}{8}(\delta_{\mu}^{\rho}\delta_{\nu}^{\sigma} + \delta_{\mu}^{\sigma}\delta_{\nu}^{\rho})\\
 & \times F_{\lambda\kappa}F_{\alpha\beta}\bar{g}^{\lambda\alpha}\bar{g}^{\kappa\beta} + 
 \frac{1}{8}\bar{g}_{\mu\nu}F_{\lambda\kappa}F_{\alpha\beta}(\bar{g}^{\lambda\rho}\bar{g}
 ^{\alpha\sigma} + \bar{g}^{\lambda\sigma}\bar{g}^{\alpha\rho})\\
 & \times \bar{g}^{\beta\kappa} + \frac{1}{8}\bar{g}_{\mu\nu}F_{\lambda\kappa}F_{\alpha\beta}
 \bar{g}^{\lambda\alpha}(\bar{g}^{\beta\rho}\bar{g}^{\kappa\sigma} + \bar{g}^{\beta\sigma}
 \bar{g}^{\kappa\rho})\\
 & - \frac{1}{2}(\delta_{\mu}^{\rho}\delta_{\nu}^{\sigma} + \delta_{\mu}^{\sigma}\delta_{\nu}
 ^{\rho})[\bar{g}^{\alpha\beta}\mathcal{D}_{\alpha}\Phi\mathcal{D}_{\beta}\Phi^{*} + V(|\Phi|)]\\
 & + \frac{1}{2}\bar{g}_{\mu\nu}(\mathcal{D}^{\rho}\Phi\mathcal{D}^{\sigma}\Phi^{*} + \mathcal{D}
 ^{\sigma}\Phi\mathcal{D}^{\rho}\Phi^{*}) - \frac{1}{2}\bar{g}^{\rho\sigma}T_{\mu\nu}.
\end{split}
\end{equation}
Considering the spontaneous symmetry breaking phenomenon, we obtain a non-zero vacuum 
expectation value of $|\Phi|$. As a result, in the leading order the above expression 
has the following component
\begin{equation}\label{S.65}
\begin{split}
\mathcal{O}_{\mu\nu}^{ \ \ \rho\sigma} & = - \frac{1}{2}(\delta_{\mu}^{\rho}\delta_{\nu}^{\sigma} 
 + \delta_{\mu}^{\sigma}\delta_{\nu}^{\rho})V(|\Phi_{0}|) + \frac{1}{2}\bar{g}^{\rho\sigma}
 \bar{g}_{\mu\nu}V(|\Phi_{0}|)\\
 & = - \frac{1}{2}V(|\Phi_{0}|)[\delta_{\mu}^{\rho}\delta_{\nu}^{\sigma}  + \delta_{\mu}^{\sigma}
 \delta_{\nu}^{\rho} - \bar{g}^{\rho\sigma}\bar{g}_{\mu\nu}],
\end{split}
\end{equation}
where $V(|\Phi_{0}|)$ turns out to be negative after spontaneous symmetry breaking of 
the field potential. Substituting all the above information including the fact $\bar{R} = 
- \kappa^{2}T$ and $\bar{R}_{\nu\sigma} = \kappa^{2}\left(T_{\nu\sigma} - \frac{1}{2}\bar{g}
_{\nu\sigma}T\right)$ where $T = - 4 V(|\Phi_{0}|)$ and $T_{\nu\sigma} = - \bar{g}_{\nu\sigma}
V(|\Phi_{0}|)$ in the leading order, we obtain the following quadratic action of spin-2 field
\begin{equation}\label{S.66}
\begin{split}
S_{h} & = \int d^{4}x\sqrt{-\bar{g}}\Big[ - \frac{1}{2}\bar{\nabla}_{\lambda}h_{\mu\nu}\bar{\nabla}
 ^{\lambda}h^{\mu\nu} + \frac{1}{2}\bar{\nabla}_{\rho}h\bar{\nabla}^{\rho}h + \bar{\nabla}^{\nu}
 h^{\rho\mu}\bar{\nabla}_{\mu}h_{\rho\nu}\\
 & - \bar{\nabla}_{\mu}h\bar{\nabla}_{\nu}h^{\mu\nu} - 2\kappa^{2}V(|\Phi_{0}|)\left(h_{\alpha\beta}h^{\alpha\beta} - \frac{1}{2}h^{2}\right)\\
 & + 2\kappa^{2}V(|\Phi_{0}|)\left(h_{\alpha\beta}h^{\alpha\beta} - \frac{1}{2}h^{2}\right)\Big]\\
 & + \kappa^{2}V(|\Phi_{0}|)\int d^{4}x\sqrt{-\bar{g}}\left(h_{\mu\nu}h^{\mu\nu} - 
 \frac{1}{2}h^{2}\right)\\
 & = \int d^{4}x\sqrt{-\bar{g}}\Big[ - \frac{1}{2}\bar{\nabla}_{\lambda}h_{\mu\nu}\bar{\nabla}
 ^{\lambda}h^{\mu\nu} + \frac{1}{2}\bar{\nabla}_{\rho}h\bar{\nabla}^{\rho}h + \bar{\nabla}^{\nu}
 h^{\rho\mu}\bar{\nabla}_{\mu}h_{\rho\nu}\\
 & - \bar{\nabla}_{\mu}h\bar{\nabla}_{\nu}h^{\mu\nu} + \kappa^{2}V(|\Phi_{0}|)\left(h_{\alpha\beta}h^{\alpha\beta} - \frac{1}{2}h^{2}\right)\Big].
\end{split}
\end{equation}

\section{Fierz-Pauli action: Standard, Generalized and Extended}
\label{appendix C}
In this section, we briefly review linearized massive gravity as described by the generalized Fierz-Pauli action~\cite{Gambuti:2021meo, Hassan:2011vm}:
\begin{equation}\label{eq. 0.1}
S_{GFP} = \int d^{D}x \Big[ \mathcal{L}_{m = 0} - \frac{1}{2}m^{2}(h_{\mu\nu}h^{\mu\nu}
 - \alpha \, h^{2})\Big] 
\end{equation}
where $m$ is the mass of spin-2 field, $h_{\mu\nu}$ is rank-2, symmetric tensor field defined about a D-dimensional Minkowski space-time $\eta_{\mu\nu} = \text{diag}(-1, 1,\ldots, 1)$, $h = \eta^{\mu\nu}h_{\mu\nu}$ is the trace of the metric perturbation,  $\alpha$ is referred to as the massive deformation parameter and $\mathcal{L}_{m = 0}$ is the Lagrangian density corresponding to the linearized  Einstein-Hilbert action, which is given by:
\begin{equation}\label{eq. 0.1a}
\begin{split}
\mathcal{L}_{m = 0}  & = - \frac{1}{2}\partial_{\lambda}h_{\mu\nu}\partial^{\lambda}
h^{\mu\nu} + \partial_{\mu}h_{\nu\lambda}\partial^{\nu}h^{\mu\lambda} - \partial_{\mu}
h^{\mu\nu}\partial_{\nu}h\\
 & + \frac{1}{2}\partial_{\lambda}h\partial^{\lambda}h \, .
\end{split}
\end{equation}
Setting $m = 0$ in the above action leads to the linearized Einstein-Hilbert action, which describes a massless helicity-2 particle (massless spin-2 field) with the gauge symmetry under the following transformation:
\begin{equation}\label{eq. 0.2}
\delta h_{\mu\nu} = \partial_{\mu}\xi_{\nu} + \partial_{\nu}\xi_{\mu},
\end{equation}
$\xi^{\mu}$ is a spacetime-dependent infinitesimal gauge transformation. In the rest of this section, we will explore implications for two values of $\alpha$: $\alpha = 1$ and $\alpha \neq 1$. Setting $\alpha = 1$ in the above action leads to the Fierz-Pauli action, while $\alpha \neq 1$ leads to dynamical $h$. We will also consider a special case $\alpha = 1/2$, which we refer to as \emph{extended Fierz-Pauli action.}  

\subsection{Fierz-Pauli action ($\alpha = 1$)}

Setting $\alpha = 1$ in the above action \eqref{eq. 0.1} and varying the action w.r.t the metric perturbation $h_{\mu\nu}$, we obtain the following equation of motion:
\begin{equation}\label{eq. 0.3}
\begin{split}
\Box h_{\mu\nu} & - \partial_{\lambda}\partial_{\mu}h_{ \ \nu}^{\lambda} - \partial_{\lambda}
\partial_{\nu}h_{ \ \mu}^{\lambda} + \eta_{\mu\nu}\partial_{\lambda}\partial_{\sigma}h^{\lambda
\sigma}\\
 & + \partial_{\mu}\partial_{\nu}h - \eta_{\mu\nu}\Box h = m^{2}(h_{\mu\nu} - \eta_{\mu\nu}h) \, .
\end{split}
\end{equation} 
These are coupled partial differential equations, and the underlying physics is not transparent. To extract the physics, we do a series of transformations: First, using the fact that the left-hand side of the above equation consists of the linearised form of the Einstein tensor whose divergence vanishes, we obtain the following relation:
\begin{equation}\label{eq. 0.4}
m^{2}(\partial^{\mu}h_{\mu\nu} - \partial_{\nu}h) = 0.
\end{equation}
Since $m \neq 0$, substituting the above result back into the equations of motion gives
the following equation
\begin{equation}\label{eq. 0.5}
\Box h_{\mu\nu} - \partial_{\mu}\partial_{\nu}h = m^{2}(h_{\mu\nu} - \eta_{\mu\nu}h).
\end{equation}
Second, taking the trace of the above expression \textit{w.r.t} the Minkowski metric we obtain $h = 0$ 
which means that the field $h_{\mu\nu}$ is traceless. Using the traceless condition and Eq.~(\ref{eq. 0.4}) leads to 
$\partial^{\mu}h_{\mu\nu} = 0$. This implies that 
$h_{\mu\nu}$ is transverse. Lastly, using transverse and traceless relations of $h_{\mu\nu}$ in Eq.~(\ref{eq. 0.5}), we get the following wave equation:
\begin{equation}\label{eq. 0.6}
(\Box - m^{2})h_{\mu\nu} = 0.
\end{equation} 
Thus, the equations of motion (\ref{eq. 0.3}) leads to a 
massive wave equation, transverse condition, and traceless constraint:
\begin{equation}\label{eq. 0.7}
(\Box - m^{2})h_{\mu\nu} = 0, \ \partial^{\mu}h_{\mu\nu} = 0, \ h = 0.
\end{equation}
Although the above three equations are exactly equivalent to (\ref{eq. 0.3}), it is 
easier to count the number of DoF by using the above set of equations. 
The first equation is the evolution of a symmetric tensor field $h_{\mu\nu}$ and implies 
$10$ DoF in $D = 4$. The traceless condition reduces one degree of freedom by forming a single constraint on the system. The transverse condition leads to $4$ more constraints. Hence, the Fierz-Pauli action consists of $5$ DoF in $D = 4$ spacetime dimension. 

One of the striking features of the massive Spin-2 field, compared to massive spins $0, 1/2, 1,$ and $3/2$ fields, is that one can not straightforwardly extend the framework from Minkowski to arbitrary curved space-times~\cite{Buchbinder:1999ar, Bernard:2014bfa, Bernard:2015uic, Mazuet:2017hey, Mazuet_2018}. As can be seen from Eq.~(\ref{S.58}), describing the massive spin-2 field in curved space-time depends on the background Ricci tensor and Ricci scalar. While the Fierz-Pauli action in Minkowski space-time has only five dynamical polarizations, it has six in arbitrary curved space-time. The 
additional polarization is a ghost~\cite{Buchbinder:1999ar, Bernard:2014bfa, Bernard:2015uic, Mazuet:2017hey, Mazuet_2018}. As can be seen from Eq.~~(\ref{S.58}), the Fierz-Pauli theory for a free massive spin-2 field can be described in Einstein spaces, defined by the relation $R_{\mu\nu} = \Lambda g_{\mu\nu}$.  

Generally, it is believed that $\alpha \neq 1$ in the massive gravity theory 
leads to ghost-like DoF with negative norm states. In our
analysis of the effect of SSB on the dynamics of the spin-2 field, we find a massive gravity
theory with extended Fierz-Pauli action where the massive deformation parameter takes the 
value $\alpha = 1/2$. In the next two sections, we have shown explicitly that $\alpha = 1/2$ 
also does not generate ghost DoF in generalized massive gravity theory.

\subsection{Generalized Fierz-Pauli action ($\alpha \neq 1$)}\label{section 3}
As mentioned above, the generalized Fierz-Pauli action \eqref{eq. 0.1} contains an arbitrary massive deformation parameter ($\alpha$). As we show, the innocuous $\alpha$ changes some of the key features of the Fierz-Pauli action. Variation of the action \eqref{eq. 0.1} w.r.t $h_{\mu\nu}$ leads to the following equations of motion:
\begin{equation}\label{eq. 0.9}
\begin{split}
\Box h_{\mu\nu} & - \partial_{\lambda}\partial_{\mu}h_{ \ \nu}^{\lambda} - \partial_{\lambda}
\partial_{\nu}h_{ \ \mu}^{\lambda} + \eta_{\mu\nu}\partial_{\lambda}\partial_{\sigma}h^{\lambda
\sigma}\\
 & + \partial_{\mu}\partial_{\nu}h - \eta_{\mu\nu}\Box h = m^{2}(h_{\mu\nu} - \alpha
 \eta_{\mu\nu}h).
\end{split}
\end{equation}
Like in the previous case, these equations are coupled partial differential equations. To extract the physics, we do a series of transformations: First, taking the partial derivative $\partial^{\mu}$ in the above expression leads to:
\begin{equation}\label{eq. 0.10}
m^{2}(\partial^{\mu}h_{\mu\nu} - \alpha\partial_{\nu}h) = 0 \, \Longrightarrow \, \partial^{\mu}h_{\mu\nu} =  \alpha\partial_{\nu}h
\end{equation}
Substituting the above relation in Eq.~(\ref{eq. 0.9}), we get:
\begin{equation}\label{eq. 0.11}
\begin{split}
\Box h_{\mu\nu} + (1 - 2\alpha)\partial_{\mu}\partial_{\nu}h + (\alpha - 1)\eta_{\mu\nu}
\Box h = m^{2}(h_{\mu\nu} - \alpha\eta_{\mu\nu}h).
\end{split}
\end{equation}
Taking the trace of the above relation, we obtain the following wave equation for the trace of $h_{\mu\nu}$:
\begin{equation}\label{eq. 0.12}
\left(\Box - \frac{(1 - \alpha D)m^{2}}{(\alpha - 1)(D - 2)}\right)h = 0.
\end{equation}
It is easy to see that if we set $D = 2$ or $\alpha = 1$ in the above expression, the mass term diverges, and $h$ becomes a non-propagating degree of freedom. Hence, the trace is not zero for $\alpha \neq 1$, and $h$ is a dynamical field. In other words, the theory has an extra degree of freedom. Thus, the generalized Fierz-Pauli action \eqref{eq. 0.1} contains six DoF.

\subsection{Extended Fierz-Pauli action ($\alpha = 1/2$) in 4-D}
 
In this subsection, we consider a specific case of 
$\alpha = \frac{1}{2}$ and $D = 4$, as the Spontaneous Symmetry Breaking of minimally coupled, self-interacting complex scalar field in 4-dimensions leads to generalized Fierz-Pauli action with $\alpha = 1/2$. Setting $\alpha = 1/2$ and $D = 4$ in the above equation \eqref{eq. 0.12} reduces to:
\begin{equation}\label{eq. 0.13}
(\Box - m^{2})h = 0.
\end{equation}
Setting $\alpha = {1}/{2}$ in Eq.~(\ref{eq. 0.11}) reduces to
\begin{equation}\label{eq. 0.14}
\Box h_{\mu\nu} - \frac{1}{2}\eta_{\mu\nu}\Box h = m^{2}\left(h_{\mu\nu} - \frac{1}{2}
\eta_{\mu\nu}h\right),
\end{equation} 
which can also be expressed in the following manner
\begin{equation}\label{eq. 0.15}
(\Box - m^{2})\bar{h}_{\mu\nu} = 0,
\end{equation}
where $\bar{h}_{\mu\nu} \equiv h_{\mu\nu} - \frac{1}{2}\eta_{\mu\nu}h$ is the trace-reversed metric perturbation. Therefore, for $\alpha = 1/2$ in 4-D spacetime, the above theory has $6$ DoF, which follows from the four constraint equations (transverse conditions)  (\ref{eq. 0.10})
\begin{equation}\label{eq. 0.16}
\partial^{\mu}\bar{h}_{\mu\nu} = 0.
\end{equation}

\section{Stueckelberg mechanism in linearised massive gravity theory}\label{Appendix D}
Massive gravity theories are often plagued by ghosts \cite{Creminelli:2005qk}. This section discusses this issue in detail for the Fierz-Pauli action ($\alpha = 1$) and extended Fierz-Pauli action ($\alpha = 1/2$). We qualitatively discuss for $\alpha \neq 1$. 

The action of linearised massive gravity theory coupled to the source term in the Minkowski background is given by:
\begin{equation}\label{S.1}
S = S_{GFP} + \int d^{D}x  \kappa h_{\mu\nu}T^{\mu\nu} \, ,
\end{equation}
where  $S_{GFP}$ is the generalized Fierz-Pauli action \eqref{eq. 0.1} and $T^{\mu\nu}$ is the energy-momentum tensor of the source. Coupling the massive gravity with external fields leads to apparent singularity in the $m \to 0$ limit. 

\subsection{Fierz-Pauli action ($\alpha = 1$)}

In this subsection, we set $\alpha = 1$ in the above action. To isolate the DoF, it is helpful to introduce the Stueckelberg fields in a series of steps \cite{Ruegg:2003ps}.  

The first step is to demand that the above action preserves the gauge symmetry --- $\delta h_{\mu\nu} = \partial_{\mu}\xi_{\nu} + \partial_{\nu}\xi_{\mu}$ --- introduces the first Stueckelberg field ($V_{\mu}$). Under the following transformation
\begin{equation}\label{S.2}
h_{\mu\nu} \rightarrow h_{\mu\nu} + \partial_{\mu}V_{\nu} + \partial_{\nu}V_{\mu},
\end{equation} 
the action (\ref{S.1}) takes the following form:
\begin{equation}\label{S.3}
\begin{split}
S & = \int d^{D}x \Big[\mathcal{L}_{m = 0} - \frac{1}{2}m^{2}(h_{\mu\nu}h^{\mu\nu} - h^{2})
- \frac{1}{2}m^{2}F_{\mu\nu}F^{\mu\nu}\\
 & - 2m^{2}(h_{\mu\nu}\partial^{\mu}V^{\nu} - h\partial_{\mu}V^{\mu}) + \kappa h_{\mu\nu}
 T^{\mu\nu} - 2\kappa V_{\mu}\partial_{\nu}T^{\mu\nu}\Big],
\end{split}
\end{equation}
where $F_{\mu\nu} = \partial_{\mu}V_{\nu} - \partial_{\nu}V_{\mu}$. 
[Note that under the transformation 
$\mathcal{L}_{m = 0}$ remains invariant; however, other terms in action (\ref{S.1}) are not invariant.] The above action now has a gauge symmetry, which is defined
by
\begin{equation}\label{S.4}
\delta h_{\mu\nu} = \partial_{\mu}\xi_{\nu} + \partial_{\nu}\xi_{\mu}, \ \delta V_{\mu} = - 
\xi_{\mu}.
\end{equation}
At this point, one might think to consider scaling $V_{\mu} \rightarrow V_{\mu}/m$ 
to normalize the vector kinetic term, then take the $m \rightarrow 0$ limit. However, in this 
scenario, the resulting massless spin-2 field and massless photon would yield 4 DoF (in 4 dimensions). Hence, $m \rightarrow 0$ is singular, leading to the loss of one of the original 5 DoF. 

The second step is to introduce another Stueckelberg field $\phi$:
\begin{equation}\label{S.5}
V_{\mu} \rightarrow V_{\mu} + \partial_{\mu}\phi.
\end{equation}
Under this transformation, the action (\ref{S.3}) becomes:
\begin{equation}\label{S.6}
\begin{split}
S & = \int d^{D}x \Big[\mathcal{L}_{m = 0} - \frac{1}{2}m^{2}(h_{\mu\nu}h^{\mu\nu} - h^{2})
- \frac{1}{2}m^{2}F_{\mu\nu}F^{\mu\nu}\\
 & - 2m^{2}(h_{\mu\nu}\partial^{\mu}V^{\nu} - h\partial_{\mu}V^{\mu}) - 2m^{2}(h_{\mu\nu}
 \partial^{\mu}\partial^{\nu}\phi - h\Box\phi)\\
 & + \kappa h_{\mu\nu}T^{\mu\nu} - 2\kappa V_{\mu}\partial_{\nu}T^{\mu\nu} + 2\kappa\phi
 \partial_{\mu}\partial_{\nu}T^{\mu\nu}\Big].
\end{split}
\end{equation}
The above action has two independent gauge symmetries:
\begin{equation}\label{S.7}
\begin{split}
\delta h_{\mu\nu} & = \partial_{\mu}\xi_{\nu} + \partial_{\nu}\xi_{\mu}, \ \delta V_{\mu} = 
- \xi_{\mu}\\
\delta V_{\mu} & = \partial_{\mu}\Lambda, \ \delta\phi = - \Lambda.
\end{split}
\end{equation}
Rescaling $V_{\mu} \rightarrow \frac{1}{m}V_{\mu}, \ \phi\rightarrow\frac{1}{m^{2}}\phi$ in the above action (\ref{S.6}) leads to:
\begin{equation}\label{S.8}
\begin{split}
S & = \int d^{D}x \Big[\mathcal{L}_{m = 0} - \frac{1}{2}m^{2}(h_{\mu\nu}h^{\mu\nu} - h^{2})
- \frac{1}{2}F_{\mu\nu}F^{\mu\nu}\\
 & - 2m(h_{\mu\nu}\partial^{\mu}V^{\nu} - h\partial_{\mu}V^{\mu}) - 2(h_{\mu\nu}
 \partial^{\mu}\partial^{\nu}\phi - h\Box\phi)\\
 & + \kappa h_{\mu\nu}T^{\mu\nu} - \frac{2}{m}\kappa V_{\mu}\partial_{\nu}T^{\mu\nu} + 
 \frac{2}{m^{2}}\kappa\phi\partial_{\mu}\partial_{\nu}T^{\mu\nu}\Big] \, .
\end{split}
\end{equation}
This action is invariant under the gauge transformations:
\begin{equation}\label{S.9}
\begin{split}
\delta h_{\mu\nu} & = \partial_{\mu}\xi_{\nu} + \partial_{\nu}\xi_{\mu}, \ \delta V_{\mu} = 
- m\xi_{\mu}\\
\delta V_{\mu} & = \partial_{\mu}\Lambda, \ \delta\phi = - m\Lambda.
\end{split}
\end{equation}

The third step is to impose conservation of energy-momentum tensor and take the limit $m \rightarrow 0$ in action (\ref{S.8}). This leads to:
\begin{equation}\label{S.10}
\begin{split}
S & = \int d^{D}x \Big[\mathcal{L}_{m = 0} - \frac{1}{2}F_{\mu\nu}F^{\mu\nu}
 - 2(h_{\mu\nu}\partial^{\mu}\partial^{\nu}\phi - h\Box\phi)\\
 & + \kappa h_{\mu\nu}T^{\mu\nu}
 \Big] \, .
\end{split}
\end{equation}
The above action represents a scalar-tensor-vector theory with massless vector and tensor fields. 
Interestingly, the vector is decoupled, but the scalar is kinetically mixed with the tensor.

The fourth step is to un-mix the scalar and tensor DoF at the expense of the minimal coupling to $T_{\mu\nu}$. The following field redefinition achieves this:
\begin{equation}\label{S.11}
h_{\mu\nu} = h_{\mu\nu}' + \pi\eta_{\mu\nu},
\end{equation}
where $\pi$ is an arbitrary scalar field. The above transformation can be interpreted as a linear version of the conformal transformation. The change in the massless spin-2 part is given by
\begin{equation}\label{S.12}
\begin{split}
\mathcal{L}_{m = 0}[h] & = \mathcal{L}_{m = 0}[h'] + (D - 2)\Big[\partial_{\mu}\pi\partial^{\mu}h'
- \partial_{\mu}\pi\partial_{\nu}h^{'\mu\nu}\\
 & + \frac{1}{2}(D - 1)\partial_{\mu}\pi\partial^{\mu}\pi\Big].
\end{split}
\end{equation} 
This is nothing but the linearisation of the effect of a conformal transformation on the 
Einstein-Hilbert action. Setting $\pi = 
\frac{2}{D - 2}\phi$, the action (\ref{S.10}) reduces to:
\begin{equation}\label{S.13}
\begin{split}
S & = \int d^{D}x \Big[\mathcal{L}_{m = 0}[h'] - \frac{1}{2}F_{\mu\nu}F^{\mu\nu} - \frac{D - 1}
{D - 2}\partial_{\mu}\phi\partial^{\mu}\phi\\
 & + \kappa h_{\mu\nu}'T^{\mu\nu} + \frac{2}{D - 2}\kappa\phi T\Big],
\end{split}
\end{equation}
where $T$ is the trace of the energy-momentum tensor. As a result, it is now easy to see that the 
gauge transformations now look like
\begin{equation}\label{S.14}
\begin{split}
\delta h_{\mu\nu} & = \partial_{\mu}\xi_{\nu} + \partial_{\nu}\xi_{\mu}, \ \delta V_{\mu} = 
0\\
\delta V_{\mu} & = \partial_{\mu}\Lambda, \ \delta\phi = 0.
\end{split}
\end{equation}
It is easy to see that in $D = 4$, the above action has  five DoF ---- two DoF carried by the canonical massless spin-2 field, two DoF carried by the canonical massless vector, and one DoF carried by the canonical massless scalar.
To confirm or infirm that these extra DoF do not introduce ghost DoF causing instabilities, we return to the action \eqref{S.6}.  

Fifth step is to apply the transformation $h_{\mu\nu} = 
h_{\mu\nu}' + \frac{2}{D - 2}\phi\eta_{\mu\nu}$ in the action \eqref{S.6}. Using the conservation of energy-momentum ($\partial_{\mu} T^{\mu\nu} = 0$), we get:
\begin{equation}\label{S.15}
\begin{split}
S & = \int d^{D}x \Big[\mathcal{L}_{m = 0}[h'] - \frac{1}{2}m^{2}(h_{\mu\nu}'h^{'\mu\nu} - 
h^{'2}) - \frac{1}{2}F_{\mu\nu}F^{\mu\nu}\\
 & + 2\frac{D - 1}{D - 2}\phi\left(\Box + \frac{D}{D - 2}m^{2}\right)\phi\\
 & - 2m \left( h_{\mu\nu}'\partial^{\mu}V^{\nu} - h'\partial_{\mu}V^{\mu}
- 2  \frac{D - 1}{D - 2} \phi\partial_{\mu}V^{\mu} \right)\\
 & + 2\frac{D - 1}{D - 2} m^{2}h'\phi + \kappa h_{\mu\nu}'T^{\mu\nu} + \frac{2}{D - 2}
 \phi T\Big].
\end{split}
\end{equation}
The above action shows that the scalar, vector, and tensor are all coupled. This translates into the following gauge symmetries of the action: 
%
\begin{equation}\label{S.16}
\begin{split}
\delta h_{\mu\nu}' & = \partial_{\mu}\xi_{\nu} + \partial_{\nu}\xi_{\mu} + \frac{2}{D - 2}m
\Lambda\eta_{\mu\nu}, \ \delta V_{\mu} = - m\xi_{\mu}\\
\delta V_{\mu} & = \partial_{\mu}\Lambda, \ \delta\phi = - m\Lambda.
\end{split}
\end{equation}
The last step is to decouple the scalar, vector, and tensor fields in the action \eqref{S.15} by removing the gauge redundancies on the vector and tensor fields. Imposing the following Lorentz-like gauge conditions
\begin{equation}\label{S.17}
\begin{split}
S_{GF1} & = - \int d^{D}x \left(\partial^{\nu}h_{\mu\nu}' - \frac{1}{2}\partial_{\mu}h'
 + mV_{\mu}\right)^{2}\\
S_{GF2} & = - \int d^{D}x \left(\partial_{\mu}V^{\mu} + m\left(\frac{1}{2}h' + 2\frac{D - 1}
{D - 2}\phi\right)\right)^{2},  
\end{split}
\end{equation}
in the above action, removes the first term in the second line of  Eq.~\eqref{S.15}: 
%
\begin{equation}\label{S.18}
\begin{split}
S & + S_{GF1} + S_{GF2} = \int d^{D}x \Big[\frac{1}{2}h_{\mu\nu}'(\Box - m^{2})h^{'\mu\nu}
- \frac{1}{4}h'(\Box - m^{2})h'\\
 & + V_{\mu}(\Box - m^{2})V^{\mu} + 2\frac{D - 1}{D - 2}\phi(\Box - m^{2})\phi + 
 \kappa h_{\mu\nu}'T^{\mu\nu} + \frac{2}{D - 2}\kappa\phi T\Big].
\end{split}
\end{equation}
From the above action, we can read the propagators of the tensor, vector, and scalar field in momentum space is
\begin{equation}\label{S.19}
\begin{split}
\frac{- i}{p^{2} + m^{2}} & \Big[\frac{1}{2}(\eta_{\alpha\sigma}\eta_{\beta\lambda} + 
\eta_{\alpha\lambda}\eta_{\beta\sigma}) - \frac{1}{D - 2}\eta_{\alpha\beta}\eta_{\sigma\lambda}
\Big], \ \frac{1}{2}\frac{-i\eta_{\mu\nu}}{p^{2} + m^{2}},\\
 & \frac{(D - 2)}{4(D - 1)}\frac{-i}{p^{2} + m^{2}}.
\end{split}
\end{equation}
This confirms that the $\alpha = 1$ theory has no ghost DoF.

\subsection{Extended Fierz-Pauli action ($\alpha = 1/2$) in 4-D}

In this subsection, we set $\alpha = 1/2$ in action \eqref{S.1} and show that the resultant theory has no ghost instability. We then discuss why the result is invalid for other values of $\alpha$ and $D$-dimensions.

The first step is to decompose the metric in terms of the helicity-2 field and spin-1 field:
\begin{equation}\label{L1}
h_{\mu\nu} = h_{\mu\nu}^{(T)} + \partial_{\mu}h_{\nu}^{(L)} + \partial_{\nu}h_{\mu}^{(L)} \,  .
\end{equation}
$\mathcal{L}_{m = 0}$ in action \eqref{S.1} remains invariant, however, the mass term becomes
\begin{equation}\label{L2}
\begin{split}
- \frac{1}{2}m^{2} & \left(h_{\mu\nu}h^{\mu\nu} - \frac{1}{2}h^{2}\right) = - \frac{1}{2}
m^{2}\Big[h_{\mu\nu}^{(T)}h^{(T)\mu\nu}\\
 & + 2h_{\mu\nu}^{(T)}(\partial^{\mu}h^{(L)\nu} + 
\partial^{\nu}h^{(L)\mu}) + 2\partial_{\mu}h_{\nu}^{(L)}\partial^{\mu}h^{(L)\nu}\\
 & + 2\partial_{\mu}h_{\nu}^{(L)}\partial^{\nu}h^{(L)\mu} - \frac{1}{2}(h^{(T)2} + 4h^{(T)}
 \partial_{\mu}h^{(L)\mu}\\
 & + 4(\partial_{\mu}h^{(L)\mu})^{2})\Big].
\end{split}
\end{equation}
The second step is to further decompose the spin-1 field into helicity-1 and a spin-0 field in the following manner:
\begin{equation}\label{L3}
h_{\mu}^{(L)} = l_{\mu}^{\perp} + \partial_{\mu}l^{\parallel}.
\end{equation}
Under the above-decomposition, the higher derivative term in $l^{\parallel}$ can be simplified as:
\begin{equation}\label{L4}
\begin{split}
- \frac{1}{2}m^{2} & (4\partial_{\mu}\partial_{\nu}l^{\parallel}\partial^{\mu}\partial^{\nu}
l^{\parallel} - 2(\Box l^{\parallel})^{2}) = - m^{2}(\Box l^{\parallel})^{2}\\
 & = - m^{2}l^{\parallel}(\Box)^{2}l^{\parallel},
\end{split}
\end{equation}
where we have ignored the boundary term. As a result, the expression in Eq.~(\ref{L2}) can be rewritten as:
\begin{equation}
\begin{split}
- \frac{1}{2}m^{2} & \left(h_{\mu\nu}h^{\mu\nu} - \frac{1}{2}h^{2}\right) = - \frac{1}{2}
m^{2}\Big[h_{\mu\nu}^{(T)}h^{(T)\mu\nu} - \frac{1}{2}h^{(T)2}\\
 & - 4h^{(L)\nu}\left(\partial^{\mu}h_{\mu\nu}^{(T)} - \frac{1}{2}\partial_{\nu}h^{(T)}\right)\\
 & + 2\partial_{\mu}l_{\nu}^{\perp}\partial^{\mu}l^{\perp\nu} - 2m^{2}l^{\parallel}(\Box)^{2}l^{\parallel}\Big]\\
 & = - \frac{1}{2}m^{2}\Big[h_{\mu\nu}^{(T)}h^{(T)\mu\nu} - \frac{1}{2}h^{(T)2} + 2\partial
 _{\mu}l_{\nu}^{\perp}\partial^{\mu}l^{\perp\nu}\\
 & + 2l^{\parallel}(\Box)^{2}l^{\parallel}\Big],
\end{split}
\end{equation}
where we used the transversality and the gauge fixing conditions $\partial^{\mu}
\bar{h}_{\mu\nu}^{(T)} = 0$. This shows we can decompose the spin-2 field into helicity-2, helicity-1, and spin-0 fields. 

The next step is to show that none of these are ghost DoF. To go about that, we rewrite the action \eqref{S.1} (with $\alpha = 1/2$) as:
\begin{equation}\label{L5}
\begin{split}
S & = \int d^{4}x [\mathcal{L}_{FP} + \mathcal{L}_{1}], \ \mathcal{L}_{1} = - \frac{1}{4}m^{2} h^{2} \\
& = \int d^{4}x \Big[ - \frac{1}{2}\partial_{\lambda}h_{\mu\nu}\partial^{\lambda}h^{\mu\nu}
+ \partial_{\mu}h_{\nu\lambda}\partial^{\nu}h^{\mu\lambda} - \partial_{\mu}h^{\mu\nu}\partial
_{\nu}h\\
 & + \frac{1}{2}\partial_{\lambda}h\partial^{\lambda}h - \frac{1}{2}m^{2}\left(h_{\mu\nu}h^{\mu\nu} - h^{2}\right) - \frac{1}{4}m^{2}h^{2}\Big] \, . 
\end{split}
\end{equation}
where $\mathcal{L}_{FP} $ corresponds to the Fierz-Pauli action. In the previous subsection, we isolated the DoF by introducing the Stueckelberg fields. In the rest of this subsection, we will concentrate on the second term $\mathcal{L}_{1}$.

Under the transformation $h_{\mu\nu} \rightarrow h_{\mu\nu} + \partial_{\mu}V_{\nu} + \partial_{\nu} V_{\mu}$, $\mathcal{L}_{1}$ transforms as:
\begin{equation}\label{L6}
\mathcal{L}_{1} = - \frac{1}{4}m^{2}\Big[h^{2} + 4h\partial_{\mu}V^{\mu} + 4(\partial_{\mu}
V^{\mu})^{2}\Big].
\end{equation}
Under the transformation $V_{\mu} \rightarrow V_{\mu} + \partial_{\mu}\phi$, the above expression transforms as
\begin{equation}\label{L7}
\begin{split}
\mathcal{L}_{1} & = - \frac{1}{4}m^{2}\Big[h^{2} + 4h\partial_{\mu}V^{\mu} + 4(\partial_{\mu}
V^{\mu})^{2} + 4h\Box\phi\\
 & + 8\partial_{\mu}V^{\mu}\Box\phi + 4(\Box\phi)^{2} \Big].
\end{split}
\end{equation}
It is interesting to note that $\mathcal{L}_{1}$, like $\mathcal{L}_{FP}$, is also invariant under the set of gauge transformations defined in Eq.~ (\ref{S.7}). Like in previous subsection, rescaling $V_{\mu} \rightarrow \frac{1}{m}V_{\mu}, \ \phi \rightarrow \frac{1}{m^{2}} \phi$ in Eq.~\eqref{L6} we get:
\begin{equation}\label{L8}
\begin{split}
\mathcal{L}_{1} & = - \frac{1}{4}m^{2}h^{2} - m h\partial_{\mu}V^{\mu} - (\partial_{\mu}
V^{\mu})^{2} - h\Box\phi\\
 & - 2\frac{1}{m}\partial_{\mu}V^{\mu}\Box\phi - \frac{1}{m^{2}}(\Box\phi)^{2}.
\end{split}
\end{equation}
Using the field redefinitions (\ref{S.11}) in the above expression, leads to:
\begin{equation}\label{L9}
\begin{split}
\mathcal{L}_{1} & = - \frac{1}{4}m^{2}h^{'2} - m h'\partial_{\mu}V^{\mu} - (\partial_{\mu}
V^{\mu})^{2} - h'\Box\phi - 2\frac{1}{m}\partial_{\mu}V^{\mu}\Box\phi\\
 & - \frac{1}{m^{2}}(\Box\phi)^{2} - 2m^{2}h'\pi - 4m^{2}\pi^{2} - 4m\pi\partial_{\mu}V^{\mu}
 - 4\pi\Box\phi.
\end{split}
\end{equation}
Like earlier, choosing $\pi = \phi$, we get:
\begin{equation}\label{L10}
\begin{split}
\mathcal{L}_{1} & = - \frac{1}{4}m^{2}h^{'2} - m h'\partial_{\mu}V^{\mu} - (\partial_{\mu}
V^{\mu})^{2} - h'\Box\phi - 2\frac{1}{m}\partial_{\mu}V^{\mu}\Box\phi\\
 & - \frac{1}{m^{2}}(\Box\phi)^{2} - 2m^{2}h'\phi - 4m^{2}\phi^{2} - 4m\phi\partial_{\mu}V^{\mu}
 - 4\phi\Box\phi.
\end{split}
\end{equation}
Therefore, the total action \eqref{L5} can be rewritten as:
\begin{equation}\label{L11}
\begin{split}
S & = \int d^{4}x \Big[\mathcal{L}_{m = 0}[h'] - \frac{1}{2}m^{2}(h_{\mu\nu}'h^{'\mu\nu} - 
h^{'2}) - \frac{1}{2}F_{\mu\nu}F^{\mu\nu}\\
 & + \left(- \phi\Box\phi + 2m^{2}\phi^{2}\right) - 2m(h_{\mu\nu}'\partial^{\mu}V^{\nu} - h'\partial_{\mu}V^{\mu})\\
 & + (m^{2}h'\phi + 2m\phi\partial_{\mu}V^{\mu}) + \kappa h_{\mu\nu}'T^{\mu\nu} + \kappa \phi T\\
 & - \frac{1}{4}m^{2}h^{'2} - m h'\partial_{\mu}V^{\mu} - (\partial_{\mu}
V^{\mu})^{2} - h'\Box\phi - 2\frac{1}{m}\partial_{\mu}V^{\mu}\Box\phi\\
 & - \frac{1}{m^{2}}(\Box\phi)^{2}\Big] \, .
\end{split}
\end{equation}
The last step is to decouple the scalar, vector, and tensor fields in the above action by removing the gauge redundancies on the vector and tensor fields. 
Imposing the following  gauge-fixing conditions
\begin{equation}\label{L12}
\begin{split}
S_{GF1} & = - \int d^{4}x \left(\partial^{\nu}h_{\mu\nu}' - \frac{1}{2}\partial_{\mu}h'
 + mV_{\mu}\right)^{2}\\
S_{GF2} & = \int d^{4}x \left(\partial_{\mu}V^{\mu} + m\left(\frac{1}{2}h' - \phi\right)
 + \frac{1}{m}\Box\phi\right)^{2},  
\end{split}
\end{equation}
in action \eqref{L11} leads to
\begin{equation}\label{L13}
\begin{split}
S & + S_{GF1} + S_{GF2} = \int d^{4}x \Big[\mathcal{L}_{m = 0}[h'] - \frac{1}{2}m^{2}
(h_{\mu\nu}'h^{'\mu\nu} - h^{'2})\\
 & - \frac{1}{2}F_{\mu\nu}F^{\mu\nu} + \left( - \phi\Box\phi + 3m^{2}\phi^{2}\right)
  - \left(\partial^{\nu}h_{\mu\nu}' - \frac{1}{2}\partial_{\mu}h'\right)^{2}\\
 & - m^{2}V_{\mu}V^{\mu} + \kappa h_{\mu\nu}'T^{\mu\nu} + \phi T\Big]\\
 & = \int d^{4}x \Big[\frac{1}{2}h_{\mu\nu}'(\Box - m^{2})h^{'\mu\nu} - \frac{1}{4}h'(\Box - 
 m^{2})h'\\
  & + V_{\mu}(\eta^{\mu\nu}\Box - \partial^{\mu}\partial^{\nu} - m^{2}\eta^{\mu\nu})V_{\nu}\\
 & - 3\phi(\Box - m^{2})\phi + \kappa h_{\mu\nu}'T^{\mu\nu} + \kappa\phi T\Big].
\end{split}
\end{equation}

The relative sign of the first and third terms in the last line of the above equation does not change the equations of motion. Therefore, classically, such a change in sign is valid. Moreover, the above alterations of two signs do not change the theory even at the quantum level since all the correlation functions in quantum field theory that come from a single generating functional remain unchanged.
The above action suggests no ghost DoF for $\alpha = 1/2$ in four dimensions. However, for an arbitrary $\alpha$ in general $D$-dimensional space-time,  the coefficients of each term may not be such that we can absorb the $(\Box\phi)^{2}$ in the gauge-fixing action.

\section{Degravitation}\label{appendix: degravitation}

As mentioned in the main text, the mass of spin-2 field can be interpreted as a high-pass filter that sources must pass through before being detected by the spin-2 field \cite{Dvali:2007kt, deRham:2014zqa, Modesto:2013jea}. This is analogous to the propagation of electromagnetic waves in plasma. In unmagnetized neutral plasma, the dispersion relation for electromagnetic waves is $\omega^2 = k^2 + \omega_p^2$, where $\omega_p$ is plasma frequency. Thus, waves with frequencies lower than \(\omega _{p}\) cannot propagate in plasma. In the case of massive gravity, the short wavelength has little effect. As the cosmological constant is inherently a constant, it operates like a source with an exceedingly long wavelength. Thus, the cosmological constant acts as a screen that serves as a high-pass filter, leading to a considerably diminished observed effective value. This process is commonly referred to as degravitation \cite{Dvali:2007kt, deRham:2014zqa, Modesto:2013jea}.

In the rest of this section, we derive this for the Fierz-Pauli action ($\alpha = 1$) and the extended Fierz-Pauli action ($\alpha = 1/2$).

\subsection{Fierz-Pauli action ($\alpha = 1$)}

As discussed in the previous two sections, the Fierz-Pauli action ($\alpha = 1$) contains 5 DoF --- two helicity-2 modes, two helicity-1 modes, and one scalar DoF. To explicitly see the degravitation mechanism, we first need to isolate the observable DoF (helicity-2 modes) from the other DoF in the action \eqref{S.1}. After this, we integrate the extra DoF from the action. 

To do this, we start with the action (\ref{S.3}) that includes one Stuekelberg vector field $A_{\mu}$:
\begin{equation}\label{S.20}
\begin{split}
S & = \int d^{D}x \Big[\mathcal{L}_{m = 0} - \frac{1}{2}m^{2}(h_{\mu\nu}h^{\mu\nu}
- h^{2}) - \frac{1}{2}m^{2}F_{\mu\nu}F^{\mu\nu}\\
 & - 2m^{2}(h_{\mu\nu}\partial^{\mu}V^{\nu} - h\partial_{\mu}V^{\mu}) + \kappa h_{\mu\nu}T^{\mu\nu}\Big].
\end{split}
\end{equation}
In the above action, $h_{\mu\nu}$ is a helicity-2 field containing the two observable DoF, while $V_{\mu}$ contains the other three.  

The second step is to substitute the generalized Lorenz-gauge condition: 
\begin{equation}\label{S.22}
\partial_{\mu}V^{\mu} = \frac{h}{2} - \mathcal{N} \, ,
\end{equation}
in the above action, leading to:
\begin{equation}\label{S.21}
\begin{split}
S & = \int d^{D}x \Bigg[\mathcal{L}_{m = 0} + m^{2}\Big[ - \frac{1}{2}h_{\mu\nu}h^{\mu\nu}
 + \frac{1}{4}h^{2} + V_{\mu}\Box V^{\mu}\\
 & + \mathcal{N}(h - \mathcal{N}) - V^{\mu}\left(\partial_{\mu}h - 2\partial^{\nu}h_{\mu\nu} + 2\partial_{\mu}\mathcal{N}\right)\Big] + \kappa h_{\mu\nu}T^{\mu\nu}\Bigg].
\end{split}
\end{equation} 
where $\mathcal{N}$ is the auxiliary scalar field. It is easy to check that substituting \eqref{S.22} in the above action leads to the action (\ref{S.20}). This shows that both the actions (\ref{S.20}) and (\ref{S.21}) are equivalent. Varying the above action w.r.t $V_{\mu}$, we get: 
\begin{equation}\label{S.23}
\Box V_{\mu} = \left(\frac{1}{2}\partial_{\mu}h - \partial^{\nu}h_{\mu\nu} + 
\partial_{\mu}\mathcal{N}\right).
\end{equation}
The third step is to rewrite the above expression using the inverse  operator ($\Box^{-1}$):
\begin{equation}\label{S.23a}
V_{\mu} = \frac{1}{\Box}\left(\frac{1}{2}\partial_{\mu}h - \partial^{\nu}h_{\mu\nu} + 
\partial_{\mu}\mathcal{N}\right).
\end{equation}
and substituting in the action (\ref{S.21}). This results in the following on-shell action
\begin{equation}\label{S.24}
\begin{split}
S & = \int d^{D}x \Big[\frac{1}{2}h_{\mu\nu}\left(1 - \frac{m^{2}}{\Box}\right)\mathcal{E}
^{\mu\nu,\alpha\beta}h_{\alpha\beta} - 2\mathcal{N}\frac{1}{\Box}(\partial_{\mu}\partial
_{\nu}h^{\mu\nu} - \Box h)\\
 & + \kappa h_{\mu\nu}T^{\mu\nu}\Big],
\end{split}
\end{equation}
where $\mathcal{E}^{\mu\nu,\alpha\beta}$ is the second order differential operator
\begin{equation}\label{S.25}
\begin{split}
\mathcal{E}^{\mu\nu,\alpha\beta} & = \left(\frac{1}{2}(\eta^{\mu\alpha}\eta^{\nu\beta} + 
\eta^{\mu\beta}\eta^{\nu\alpha}) - \eta^{\mu\nu}\eta^{\alpha\beta}\right)\Box\\
 & - 2\partial^{(\mu}\partial^{(\alpha}\eta^{ \ \beta)\nu)} + \eta^{\alpha\beta}\partial^{\mu}
\partial^{\nu} + \eta^{\mu\nu}\partial^{\alpha}\partial^{\beta}.
\end{split}
\end{equation}
The fourth step is to use the following linearized conformal transformation
\begin{equation}\label{S.26}
h_{\mu\nu} = h_{\mu\nu}' + \frac{2}{D - 2}\frac{1}{\Box - m^{2}}\mathcal{N}\eta_{\mu\nu},
\end{equation}
the action (\ref{S.24}) reduces to:
\begin{equation}\label{S.27}
\begin{split}
S & = \int d^{D}x \Big[\frac{1}{2}h_{\mu\nu}'\left(1 - \frac{m^{2}}{\Box}\right)\mathcal{E}
^{\mu\nu,\alpha\beta}h_{\alpha\beta}'\\
 & + 2\frac{D - 1}{D - 2}\mathcal{N}\frac{1}{\Box - m^{2}}
\mathcal{N} + \kappa h_{\mu\nu}'T^{\mu\nu} + \frac{2}{D - 2}\frac{1}{\Box - m^{2}}\mathcal{N}
T\Big].
\end{split}
\end{equation}
The last step is to substitute the following redefined auxiliary field variable
\begin{equation}\label{S.28}
\mathcal{N}' = \frac{1}{\Box - m^{2}}\mathcal{N},
\end{equation}
in the action (\ref{S.27}). This leads to:
\begin{equation}\label{S.29}
\begin{split}
S & = \int d^{D}x \Big[\frac{1}{2}h_{\mu\nu}'\left(1 - \frac{m^{2}}{\Box}\right)\mathcal{E}
^{\mu\nu,\alpha\beta}h_{\alpha\beta}'\\
 & + 2\frac{D - 1}{D - 2}\mathcal{N}'(\Box - m^{2})
\mathcal{N}' + \kappa h_{\mu\nu}'T^{\mu\nu} + \frac{2}{D - 2}\kappa\mathcal{N}'T\Big].
\end{split}
\end{equation}

It is crucial here to note that the equation of motion of metric perturbations 
$h_{\mu\nu}'$ in Fourier space can be expressed in the following form
\begin{equation}\label{S.29b}
\mathcal{E}^{\mu\nu,\alpha\beta}(k)h_{\alpha\beta}'(k) = \kappa \frac{k^{2}}{k^{2} + m^{2}} 
T^{\mu\nu}(k),
\end{equation}
which shows that the presence of a mass in the theory effectively reduces the strength of 
the metric perturbations at low energies. The left-hand side of the above equation 
is the same as that of massless spin-2 theory (GR). This shows clearly that a massive spin-2 
modes is equivalent to a filtered graviton coupled to the energy-momentum tensor $T_{\mu\nu}$ 
and a scalar with mass $m$ coupled with gravitational strength to the trace of energy-momentum 
tensor $T$. The scalar is the longitudinal mode responsible for the vDVZ discontinuity \cite{Hinterbichler:2011tt}.

\subsection{Extended Fierz-Pauli action ($\alpha = 1/2$) in 4-D}

As the previous two sections discussed, the extended Fierz-Pauli action ($\alpha = 1/2$) contains 6 DoF --- two helicity-2 modes, helicity-1 modes, and scalar DoF. To explicitly see the degravitation mechanism, we first need to isolate the observable DoF (helicity-2 modes) from the other DoF in the action \eqref{S.1}. After this, we integrate the extra DoF from the action. 

To go about this, we start with the action \eqref{L5} and (\ref{L6}) that includes only the gauge field $V_{\mu}$:
\begin{equation}
\begin{split}
S & = \int d^{4}x \Bigg[\mathcal{L}_{m = 0} - \frac{1}{2}m^{2}(h_{\mu\nu}h^{\mu\nu} - h^{2})
 - \frac{1}{2}m^{2}F_{\mu\nu}F^{\mu\nu}\\
 & - 2m^{2}(h_{\mu\nu}\partial^{\mu}V^{\nu} - h\partial_{\mu}V^{\mu}) + \kappa h_{\mu\nu}T^{\mu\nu}\\
 & - \frac{1}{4}m^{2}\Big[h^{2} + 4h\partial_{\mu}V^{\mu} + 
 4(\partial_{\mu}V^{\mu})^{2}\Big]\Bigg]\\
 & = \int d^{4}x \Bigg[\mathcal{L}_{m = 0} - \frac{1}{2}m^{2}\left(h_{\mu\nu}h^{\mu\nu} - 
 \frac{1}{2}h^{2}\right) + m^{2}V_{\mu}\Box V^{\mu}\\
 & - 2m^{2}\left(h_{\mu\nu}\partial^{\mu}V^{\nu} - \frac{1}{2}h\partial_{\mu}V^{\mu}\right) 
 + \kappa h_{\mu\nu}T^{\mu\nu}\Bigg]\\
 & = \int d^{4}x \Bigg[\mathcal{L}_{m = 0} - \frac{1}{2}m^{2}\left(h_{\mu\nu}h^{\mu\nu} - 
 \frac{1}{2}h^{2}\right) + m^{2}V_{\mu}\Box V^{\mu}\\
 & + 2m^{2}V^{\mu}\left(\partial^{\nu}h_{\mu\nu} - \frac{1}{2}\partial_{\mu}h\right)
  + \kappa h_{\mu\nu}T^{\mu\nu}\Bigg].
\end{split}
\end{equation}
Varying the above action w.r.t $V_{\mu}$, we get: 
\begin{equation}\label{S.23b}
\Box V_{\mu} = \left(\frac{1}{2}\partial_{\mu}h - \partial^{\nu}h_{\mu\nu} \right).
\end{equation}
The second step is to rewrite the above expression using the inverse  operator ($\Box^{-1}$):
\begin{equation}
V_{\mu} = - \frac{1}{\Box}\left(\partial^{\nu}h_{\mu\nu} - \frac{1}{2}\partial_{\mu}h\right),
\end{equation}
Moreover, substituting in the above action. This results in the following
on-shell action
\begin{equation}
\begin{split}
S & = \int d^{4}x \Bigg[\mathcal{L}_{m = 0} - \frac{1}{2}m^{2}\left(h_{\mu\nu}h^{\mu\nu} 
 - \frac{1}{2}h^{2}\right)\\
 & - \left(\partial^{\nu}h_{\mu\nu} - \frac{1}{2}\partial_{\mu}h
 \right)\frac{m^{2}}{\Box}\left(\partial_{\rho}h^{\mu\rho} - \frac{1}{2}\partial^{\mu}h
 \right) + \kappa h_{\mu\nu}T^{\mu\nu}\Bigg]\\
 & = \int d^{4}x \Bigg[- \frac{1}{2}\partial_{\lambda}\bar{h}_{\mu\nu}\partial^{\lambda}
 \bar{h}^{\mu\nu} + \partial_{\mu}\bar{h}_{\nu\lambda}\partial^{\nu}\bar{h}^{\mu\lambda} 
 + \frac{1}{4}\partial_{\lambda}\bar{h}\partial^{\lambda}\bar{h}\\
 & - \frac{1}{2}m^{2}\left(
 \bar{h}_{\mu\nu}\bar{h}^{\mu\nu} - \frac{1}{2}\bar{h}^{2}\right) - \partial^{\nu}\bar{h}
 _{\mu\nu}\frac{m^{2}}{\Box}\partial_{\rho}\bar{h}^{\mu\rho}\\
 & + \kappa \bar{h}_{\mu\nu}T^{\mu\nu} - \frac{\kappa}{2}\bar{h}T\Bigg]\\
 & = \int d^{4}x \Bigg[\frac{1}{2}\bar{h}_{\mu\nu}\left(1 - \frac{m^{2}}{\Box}\right)
 \bar{\mathcal{E}}^{\mu\nu,\alpha\beta}\bar{h}_{\alpha\beta} + \kappa \bar{h}_{\mu\nu}
 T^{\mu\nu} - \frac{\kappa}{2}\bar{h}T\Bigg],
\end{split}
\end{equation}
where
\begin{equation}
\bar{\mathcal{E}}^{\mu\nu,\alpha\beta} = \Big[\frac{1}{2}(\eta^{\mu\alpha}\eta^{\nu\beta} 
 + \eta^{\mu\beta}\eta^{\nu\alpha} - \eta^{\mu\nu}\eta^{\alpha\beta})\Box - 2\partial^{(\mu}\partial^{(\alpha}\eta^{\nu)\beta)}\Big].
\end{equation}
This shows clearly that a massive spin-2 theory for $\alpha = 1/2$ case is equivalent to a filtered spin-2 coupled to the energy-momentum tensor $T_{\mu\nu}$ and a scalar with mass $m$ coupled 
with gravitational strength to the trace of energy-momentum tensor $T$. This follows from the equation of motion of metric perturbations, which is similar to the equation (\ref{S.29b}).

\section{Hamiltonian analysis and counting of degrees of freedom}\label{appendix: Hamiltonian analysis}
\subsection{Linearised massless gravity}
First we start with the massless case as a reference. For the sake of mathematical simplicity, 
we consider the metric perturbation \textit{w.r.t} the Minkowski background in which the Lagrangian
density can be expressed as
\begin{equation}\label{E.1}
\mathcal{L}_{m = 0} = - \frac{1}{2}\partial_{\lambda}h_{\mu\nu}\partial^{\lambda}h^{\mu\nu}
+ \partial_{\mu}h_{\nu\lambda}\partial^{\lambda}h^{\mu\nu} - \partial_{\mu}h^{\mu\nu}
\partial_{\nu}h + \frac{1}{2}\partial_{\lambda}h\partial^{\lambda}h,
\end{equation}
which can be explicitly expressed as follows
\begin{equation}\label{E.2}
\begin{split}
\mathcal{L}_{m = 0} & = \frac{1}{2}(h_{jk,0})^{2} + (h_{0k,j})^{2} - \frac{1}{2}(h_{jk,l})^{2}
- h_{0k,j}h_{0j,k} - 2h_{0j,k}h_{jk,0} + h_{kj,l}h_{lj,k}\\
 & + h_{0k,k}h_{jj,0} - \frac{1}{2}h_{jj,0}h_{kk,0} - h_{00,l}h_{jj,l} + \frac{1}{2}h_{jj,l}
 h_{kk,l} + h_{0j,j}h_{kk,0}\\
 & + h_{jk,j}h_{00,k} - h_{jk,j}h_{ll,k}.
\end{split}
\end{equation}
Now we compute the conjugate momenta
\begin{equation}\label{E.3}
\begin{split}
\Pi^{00} & = \frac{\partial\mathcal{L}}{\partial h_{00,0}} = 0\\
\Pi^{0i} & = \frac{\partial\mathcal{L}}{\partial h_{0i,0}} = 0\\
\Pi^{jk} & = \frac{\partial\mathcal{L}}{\partial h_{jk,0}} = h_{jk,0} - (h_{0k,j} + h_{0j,k})
+ \delta_{jk}(2h_{0l,l} - h_{ll,0}).
\end{split}
\end{equation}
Then we can rearrange to see the following relation
\begin{equation}\label{E.4}
h_{jk,0} = \Pi_{jk} + (h_{0k,j} + h_{0j,k}) - \delta_{jk}(2h_{0l,l} - h_{ll,0}).
\end{equation}
Taking the trace of $\Pi^{jk}$, we obtain the following relation
\begin{equation}\label{E.5}
\Pi_{ll} = - 2h_{ll,0} + 4h_{0l,l} \implies (2h_{0l,l} - h_{ll,0}) = \frac{1}{2}\Pi_{ll}.
\end{equation}
Using the above relations, we may express the conjugate momenta as
\begin{equation}\label{E.6}
\begin{split}
\Pi_{jk} & = h_{jk,0} - (h_{0k,j} + h_{0j,k}) + \frac{1}{2}\delta_{jk}\Pi_{ll}\\
\implies h_{jk,0} & = \Pi_{jk} + (h_{0k,j} + h_{0j,k}) - \frac{1}{2}\delta_{jk}\Pi_{ll}.
\end{split}
\end{equation}
As a result, the Hamiltonian can be expressed as
\begin{equation}\label{E.7}
\mathcal{H}_{0} = \Pi^{\mu}h_{\mu\nu,0} - \mathcal{L} = \Pi^{jk}h_{jk,0} - \mathcal{L}.
\end{equation}
After substituting the previous results, we obtain the following expression for the 
Hamiltonian density
\begin{equation}\label{E.8}
\begin{split}
\mathcal{H}_{0} & = \frac{1}{2}(\Pi_{jk})^{2} - \frac{1}{4}(\Pi_{ll})^{2} + 2h_{0k,j}\Pi_{jk}
 + \frac{1}{2}(h_{jk,l})^{2} - h_{jk,l}h_{lj,k} + h_{00,l}h_{kk,l}\\
 & - \frac{1}{2}h_{jj,l}h_{kk,l} - h_{jk,j}h_{00,k} + h_{jk,j}h_{ll,k}\\
 & \equiv \frac{1}{2}(\Pi_{jk})^{2} - \frac{1}{4}(\Pi_{ll})^{2} + 2h_{0k,j}\Pi_{jk}
 + \frac{1}{2}(h_{jk,l})^{2} - h_{jk,l}h_{lj,k} - h_{00}\nabla^{2}h_{kk}\\
 & - \frac{1}{2}h_{jj,l}h_{kk,l} + h_{00}h_{jk,jk} + h_{jk,j}h_{ll,k},
\end{split}
\end{equation}
where in the second line, we have done integration by parts and ignore the surface terms.
Note in the above expression of the Hamiltonian density, there is no time derivative of 
$h_{00}$ which means it acts as a Lagrange multiplier which leads to the following constraint
\begin{equation}\label{E.9}
\mathcal{C}_{1} = \nabla^{2}h_{ii} - \partial_{i}\partial_{j}h_{ij} = 0.
\end{equation} 
In the similar manner, there is no time derivative of $h_{0i}$ in the expression (\ref{E.8})
which leads to the second constraint
\begin{equation}\label{E.10}
\mathcal{C}_{2} = \partial_{j}\Pi_{ij} = 0.
\end{equation}
Moreover, it can be checked quite easily that the Poisson bracket between the above two
constraints vanishes $\{\mathcal{C}_{1}, \mathcal{C}_{2}\} = 0$.  These constraints having 
zero Poisson bracket with all the other constraints in the theory are the first class
constraints.  This essentially means that the Hamiltonian in (\ref{E.8}) is first class. 
Because of their symmetric nature, both $h_{ij}$ and $\Pi^{ij}$ have 6 independent components 
in $D = 4$. Thus, together they span a $12$ dimensional phase space. Eliminating $4$ constraints 
implies an $8$ dimensional constraint surface. Finally, by subtracting the dimension of gauge 
orbits (which is again $4$, since four constraints generate four gauge invariances) we are 
left with a total $4$ degrees of freedom. As expected, two of them are the polarization states 
of the massless graviton and other two belong to their conjugate momenta.

\subsection{Massive Fierz-Pauli action}
Here we start with the generalized Fierz-Pauli action with the following form
\begin{equation}\label{E.11}
\mathcal{L}_{FP} = \mathcal{L}_{m = 0} - \frac{1}{2}m^{2}(h_{\mu\nu}h^{\mu\nu} - \alpha h^{2}),
\end{equation}
which can further be explicitly expressed as
\begin{equation}\label{E.12}
\begin{split}
\mathcal{L}_{FP} & = \frac{1}{2}(h_{jk,0})^{2} + (h_{0k,j})^{2} - \frac{1}{2}(h_{jk,l})^{2}
- h_{0k,j}h_{0j,k} - 2h_{0j,k}h_{jk,0} + h_{kj,l}h_{lj,k}\\
 & + h_{0k,k}h_{jj,0} - \frac{1}{2}h_{jj,0}h_{kk,0} - h_{00,l}h_{jj,l} + \frac{1}{2}h_{jj,l}
 h_{kk,l} + h_{0j,j}h_{kk,0}\\
 & + h_{jk,j}h_{00,k} - h_{jk,j}h_{ll,k} - \frac{1}{2}m^{2}[(1 - \alpha)h_{00}^{2} - 2h_{0j}^{2}
 + h_{jk}^{2} + 2\alpha h_{00}h_{ll} - \alpha h_{ll}^{2}]. 
\end{split}
\end{equation}
The conjugate momenta for the massive Lagrangian are the same as for the massless Lagrangian 
since the mass term does not contain any $\partial_{0}$ terms. This leads to the following 
Hamiltonian density
\begin{equation}\label{E.13}
\begin{split}
\mathcal{H} & = \frac{1}{2}(\Pi_{jk})^{2} - \frac{1}{4}(\Pi_{ll})^{2} + 2h_{0k,j}\Pi_{jk}
 + \frac{1}{2}(h_{jk,l})^{2} - h_{jk,l}h_{lj,k} + h_{00,l}h_{kk,l}\\
 & - \frac{1}{2}h_{jj,l}h_{kk,l} - h_{jk,j}h_{00,k} + h_{jk,j}h_{ll,k} + \frac{1}{2}m^{2}
 [(1 - \alpha)h_{00}^{2} - 2h_{0j}^{2} + h_{jk}^{2}\\
 & + 2\alpha h_{00}h_{ll} - \alpha h_{ll}^{2}].
\end{split}
\end{equation}
Note that for $\alpha = 1$, $h_{0i}$ does no longer act as a Lagrange multiplier as the 
Hamiltonian density contains the quadratic term of $h_{0i}$, however, $h_{00}$ still acts 
as a Lagrange multiplier, which leads to the following constraint
\begin{equation}\label{E.14}
\mathcal{C} = \nabla^{2}h_{ii} - \partial_{i}\partial_{j}h_{ij} - m^{2}h_{ii} = 0.
\end{equation}
However, this time, a secondary constraint arises from the Poisson bracket $\{H, \mathcal{C}\}$
\begin{equation}\label{E.15}
\mathcal{C}' = \partial_{i}\partial_{j}\Pi_{ij} + \frac{1}{2}m^{2}\Pi_{ii}. 
\end{equation} 
This means the Hamiltonian (\ref{E.13}) is not a first class Hamiltonian anymore. The 
resulting set of two constraints is second class and that removes the gauge freedom 
from the theory. Therefore, the theory now possesses $10$ degrees of freedom, obtained 
by subtracting $2$ constraints from $12$ dimensional phase space, in $D = 4$. Half of 
them are the $5$ polarizations of the massive graviton and other $5$ belong to their 
conjugate momenta. On the other hand, for $\alpha = \frac{1}{2}$ (extended Fierz-Pauli
action), both $h_{00}, h_{0i}$ do not act as Lagrange multiplier since the Hamiltonian
under that condition does also contain quadratic power of $h_{00}, h_{0i}$. As a result,
there are no constraint in this case, hence, the number of total degrees of freedom is
$12$ out of which $6$ are the polarizations of the massive graviton and other $6$ belong 
to their conjugate momenta. Now out of the $6$ polarizations, one represents massive 
spin-2 field and other is the scalar dynamical degree of freedom, namely, $h$ the trace
of the spin-2 field. Moreover, earlier, we have shown that it is not a ghost degree of 
freedom both using the field equations of motion and Stueckelberg trick. This shows that
our previous statement regarding the number of degrees of freedom in the theory are indeed
correct.
As can be seen in this analysis, $\alpha = 1$ reduces one extra degree of 
freedom compared to $\alpha = 1/2$. Since the treatment is covariant and does not introduce 
any Lorenz-violating fields, $\alpha = 1/2$ does not lead to local Lorenz breaking.


\bibliographystyle{apsrev4-2}
\bibliography{Ver-GRG}

\end{document}